%% file: mvm.tex
\documentclass[aps,superscriptaddress,floatfix,nofootinbib,showpacs,amsmath,amssymb,altaffilletter,floatfix,11pt,linenumbers,notitlepage]{revtex4-1}

\input{mvm.def.tex}

\renewcommand{\thepage}{}

\renewcommand{\thepage}{\arabic{page}}
\definecolor{oldlace}{rgb}{0.99, 0.96, 0.90}
\makeatletter
\def\p@subsection{}
\def\p@subsubsection{}
\def\input@path{{sections/}{src/}{figures/ISO/figurelatex/}}
\AtBeginDocument{\let\LS@rot\@undefined}
\makeatother
\usepackage{float}
\usepackage{url}
\usepackage{subcaption}
\usepackage{grffile}

\graphicspath{{figures/TestComp/}{figures/SecondApril/}{figures/ISO/pictures/}{figures/ISO/summary/}{figures/}{src/}{figures/device/}{figures/assisted/}}

\begin{document}
\nolinenumbers
\setlength{\parindent}{1em}
\setdefaultleftmargin{1em}{1em}{}{}{}{}
\setcounter{page}{0}\thispagestyle{empty}

\title{Mechanical Ventilator Milano (MVM):\protect\\
A Novel Mechanical Ventilator Designed for Mass Production in Response to the COVID-19 Pandemic}
\begin{center}
 \includegraphics[width=0.2\textwidth]{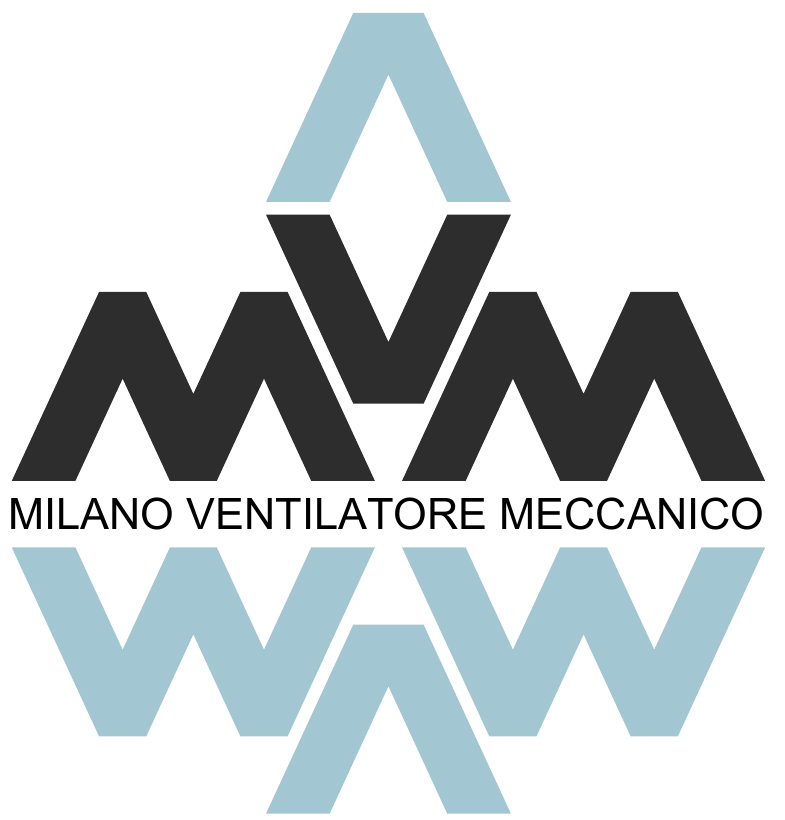}
\end{center}
\input{mvm.auth.tex}
\input{mvm.inst.tex}
\begin{abstract}
Presented here is the design of the Mechanical Ventilator Milano (MVM), a novel mechanical ventilator designed for rapid mass production in response to the \Covid\ pandemic to address the urgent shortage of intensive therapy ventilators in many countries, and the growing difficulty in procuring these devices through normal supply chains across borders. 

This ventilator is an electro-mechanical equivalent of the old and reliable Manley Ventilator, and is able to operate in both pressure-controlled and pressure-supported ventilation modes.  

MVM is optimized for the COVID-19 emergency, thanks to the collaboration with medical doctors in the front line. MVM is designed for large-scale production in a short amount of time and at a limited cost, as it relays on off-the-shelf components,  readily available worldwide

Operation of the MVM requires only a source of compressed oxygen (or compressed medical air) and electrical power.   Initial tests of a prototype device with a breathing simulator  are also presented. Further tests and developments are underway.  At this stage the MVM is not yet a certified medical device but certification is in progress.
\end{abstract}
{\let\clearpage\relax\maketitle}
\pagenumbering{arabic}   
\input{Introduction.tex}
\input{Design.tex}
\input{Controls.tex}
\input{Electronics.tex}
\input{TestComp.tex}

\input{TestISO.tex}
\input{Assisted.tex}
\input{addtional.tex}
\input{License.tex}
\bibliographystyle{mvm}
\bibliography{mvm}
\end{document}

%% file: mvm.def.tex
\usepackage[colorlinks=true,linkcolor=blue,citecolor=blue,urlcolor=blue]{hyperref}
\usepackage[separate-uncertainty,retain-explicit-plus,per-mode=symbol,binary-units]{siunitx}
\usepackage[export]{adjustbox}
\usepackage{wrapfig}
\usepackage{ragged2e}
\usepackage{array,mathtools,dcolumn}
\usepackage{amsmath,amssymb}
\usepackage{wasysym}
\usepackage{stmaryrd}
\usepackage[below]{placeins}
\usepackage[table,xcdraw]{xcolor}
\usepackage{soul}
\usepackage{tikz}
\usepackage{afterpage}
\usepackage{lineno}
\usepackage{listings}
\usepackage{array}
\usepackage[normalem]{ulem}
\useunder{\uline}{\ul}{}
\makeatletter
\@ifundefined{c@rownum}{%
  \let\c@rownum\rownum
}{}
\@ifundefined{therownum}{%
  \def\therownum{\@arabic\rownum}%
}{}
\makeatother
	\newcounter{rowno}
\usepackage[version=4]{mhchem}
\usepackage{multirow}
\usepackage{eurosym}
\usepackage{pagecolor}
\usepackage{fancyhdr}
\usepackage{etoolbox}
\usepackage{calc}
\usepackage{dcolumn}

\usepackage{bm}
\usepackage{xargs}
\usepackage{xfrac}
\newcommandx{\unsure}[2][1=]{\todo[linecolor=red,backgroundcolor=red!25,bordercolor=red,#1]{#2}}
\newcommandx{\change}[2][1=]{\todo[linecolor=blue,backgroundcolor=blue!25,bordercolor=blue,#1]{#2}}
\newcommandx{\info}[2][1=]{\todo[linecolor=OliveGreen,backgroundcolor=OliveGreen!25,bordercolor=OliveGreen,#1]{#2}}
\newcommandx{\improvement}[2][1=]{\todo[linecolor=Plum,backgroundcolor=Plum!25,bordercolor=Plum,#1]{#2}}
\newcommandx{\thiswillnotshow}[2][1=]{\todo[disable,#1]{#2}}

\newcommand{\reffig}[1]{Fig.~\ref{fig:#1}}
\newcommand{\reftab}[1]{Table~\ref{tab:#1}}

\newcommand{\refsec}[1]{Sec.~\ref{sec:#1}}

\usepackage{paralist}
\usepackage{tcolorbox}
\usepackage{pgfornament}
\usepackage{hhline}
\usepackage{pdflscape}
\usepackage{array}
\usepackage[nodayofweek]{datetime}
\usepackage{pdfpages}
\usepackage{import}
\usepackage{appendix}
\usepackage{cleveref}
\usepackage{ellipsis}
\usepackage{enumitem}
\setdescription{itemsep=0pt,parsep=0pt,labelsep=.2cm,leftmargin=2.2cm,labelwidth=2cm,listparindent=.7cm}
\setlist{nolistsep,leftmargin=1cm}
\newlist{enumcompactitem}{itemize}{3}
\setlist[enumcompactitem]{topsep=0pt,partopsep=0pt,itemsep=0pt,parsep=0pt}
\setlist[enumcompactitem,1]{label=\textbullet}
\setlist[enumcompactitem,2]{label=---}
\setlist[enumcompactitem,3]{label=*}
\newlist{enumcompactdesc}{description}{3}
\setlist[enumcompactdesc]{topsep=0pt,partopsep=0pt,itemsep=0pt,parsep=0pt}
\newlist{enumcompactenum}{enumerate}{3}
\setlist[enumcompactenum]{topsep=0pt,partopsep=0pt,itemsep=0pt,parsep=0pt}
\setlist[enumcompactenum,1]{label=\arabic*}
\setlist[enumcompactenum,2]{label=\alph*}
\setlist[enumcompactenum,3]{label=\roman*}
\newcommand{\ALSFiveK}{\mbox{\tt ALS 5000}}
\newcommand{\IngMar}{\mbox{\tt IngMar Medical}}
\newcommand{\ServoNineC}{\mbox{\tt Servo 900C}}
\newcommand{\Siemens}{\mbox{\tt Siemens}}

\DeclareSIUnit\c{\mbox{$c$}}
\DeclareSIUnit\magn{\mbox{$\times$}}
\DeclareSIUnit\min{min}
\DeclareSIUnit\hr{hr}
\DeclareSIUnit\hrs{hrs}
\DeclareSIUnit\week{week}
\DeclareSIUnit\month{mo}
\DeclareSIUnit\months{mos}
\DeclareSIUnit\year{yr}
\DeclareSIUnit\years{years}
\DeclareSIUnit\yr{yr}
\DeclareSIUnit\standard{std}
\DeclareSIUnit\str{sr}
\DeclareSIUnit\ppm{ppm}
\DeclareSIUnit\ppb{ppb}
\DeclareSIUnit\ppt{ppt}
\DeclareSIUnit\pe{PE}
\DeclareSIUnit\spe{SPE}
\DeclareSIUnit\pdm{PDM}
\DeclareSIUnit\ev{events}
\DeclareSIUnit\ct{counts}
\DeclareSIUnit\neutron{\mbox{$n$}}
\DeclareSIUnit\smp{samples}
\DeclareSIUnit\Sample{S}
\DeclareSIUnit\ch{ch}
\DeclareSIUnit\hit{hit}
\DeclareSIUnit\hits{hits}
\DeclareSIUnit\bin{(\mbox{5-PE}~bin)}
\DeclareSIUnit\sgm{\mbox{$\sigma$}}
\DeclareSIUnit\rms{RMS}
\DeclareSIUnit\keVee{\mbox{keV$_{e{\rm e}}$}}
\DeclareSIUnit\keVr{\mbox{keV$_{\rm nr}$}}
\DeclareSIUnit\eVee{\mbox{eV$_{\rm ee}$}}
\DeclareSIUnit\eVr{\mbox{eV$_{\rm nr}$}}
\DeclareSIUnit\ph{photon}
\DeclareSIUnit\el{\mbox{$e^-$}}
\DeclareSIUnit\pm{\mbox{PMT}}
\DeclareSIUnit\pixel{\mbox{pixel}}
\DeclareSIUnit\inch{''}
\DeclareSIUnit\foot{'}
\DeclareSIUnit\bit{bit}
\DeclareSIUnit\sample{samples}
\DeclareSIUnit\barn{barn}
\DeclareSIUnit\bara{bar}
\DeclareSIUnit\bar{bar}
\DeclareSIUnit\barg{barg}
\DeclareSIUnit\mlardepth{\mbox(meter~of~\LAr~depth)}
\DeclareSIUnit\Curie{Ci}
\DeclareSIUnit\PSI{psi}
\DeclareSIUnit\psia{psia}
\DeclareSIUnit\atm{atm}
\DeclareSIUnit\psf{psf}
\DeclareSIUnit\pcf{pcf}
\DeclareSIUnit\parsec{pc}
\DeclareSIUnit\cps{cps}
\DeclareSIUnit\slpm{slpm}
\DeclareSIUnit\rpm{rpm}
\DeclareSIUnit\mwe{\mbox{m.w.e.}}
\DeclareSIUnit\liveday{\mbox{live-days}}
\DeclareSIUnit\days{\mbox{days}}
\DeclareSIUnit\miles{\mbox{miles}}
\DeclareSIUnit\lumens{\mbox{lm}}
\DeclareSIUnit\degreeC{\mbox{$^{\circ}$C}}
\DeclareSIUnit\degreeF{\mbox{$^{\circ}$F}}
\DeclareSIUnit\electron{\mbox{$e^-$}}
\DeclareSIUnit\Euro{\mbox{\euro}}
\DeclareSIUnit\cph{cph}
\DeclareSIUnit\neq{neq}
\DeclareSIUnit\normal{\mbox{N}}
\DeclareSIUnit\USD{\mbox{\$}}
\DeclareSIUnit\cmw{\cm\,\ce{H2O}}
\newcommand{\MVM}{\mbox{MVM}}
\newcommand{\PCV}{\mbox{PCV}}
\newcommand{\PSV}{\mbox{PSV}}

\newcommand{\PEEP}{\mbox{PEEP}}
\newcommand{\FiOTwo}{\mbox{FiO2}}
\newcommand{\Covid}{\mbox{COVID-19}}
\newcommand{\CovidPath}{{SARS-CoV-2}}

\newcommand{\NC}{\mbox{NC}}
\newcommand{\EMC}{\mbox{EMC}}
\newcommand{\GUI}{\mbox{GUI}}
\newcommand{\MVMLicense}{\mbox{\tt CERN-OHL-S v2}}
\newcommand{\SFOne}{\mbox{SF-1}}
\newcommand{\SFTwo}{\mbox{SF-2}}
\newcommand{\GBOne}{\mbox{GB-1}}
\newcommand{\PVOne}{\mbox{PV-1}}
\newcommand{\PVTwo}{\mbox{PV-2}}
\newcommand{\PVThree}{\mbox{PV-3}}
\newcommand{\PVFour}{\mbox{PV-4}}
\newcommand{\PVFive}{\mbox{PV-5}}
\newcommand{\PVSix}{\mbox{PV-6}}
\newcommand{\OSOne}{\mbox{OS-1}}
\newcommand{\PSOne}{\mbox{PS-1}}
\newcommand{\PSTwo}{\mbox{PS-2}}
\newcommand{\PSThree}{\mbox{PS-3}}
\newcommand{\PSFour}{\mbox{PS-4}}

\newcommand{\SPOne}{\mbox{SP-1}}
\newcommand{\SPTwo}{\mbox{SP-2}}
\newcommand{\CTOne}{\mbox{CT-1}}
\newcommand{\SMOne}{\mbox{SM-1}}

\newcommand{\APLPressureRange}{\SIrange{20}{80}{\cmw}}
\newcommand{\PEEPPressureRange}{\SIrange{5}{20}{\cmw}}
\newcommand{\PEEPPressure}{\SI{5}{\cmw}}

\newcommand{\MaskTubeSize}{\SI{22}{\mm}}

\newcommand{\PSOneDiffPressure}{\SI{60}{\cmw}}
\newcommand{\PSOneResolution}{\SI{0.2}{\cmw}}
\newcommand{\SupplyPower}{\SI{50}{\watt}}
\newcommand{\BatteryLifetime}{\SI{120}{\minute}}
\newcommand{\BatteryVoltage}{\SI{12}{\volt}}
\newcommand{\BatteryCharge}{\SI{1.2}{\ampere\hour}}

\newcommand{\RespiratoryRateSedatedRange}{\SIrange{4}{50}{\rpm}}
\newcommand{\RespiratoryRateSedatedStep}{\SI{1}{\rpm}}
\newcommand{\InspiratoryTimeSedatedRange}{\SIrange{0.4}{1.5}{\second}}
\newcommand{\InspiratoryTimeSedatedStep}{\SI{0.1}{\second}}
\newcommand{\PEEPSedatedRange}{\SIrange{5}{20}{\cmw}} 
\newcommand{\InspiratoryPressureMaxSedatedRange}{\SIrange{20}{80}{\cmw}}
\newcommand{\FiOTwoSedatedRange}{\SIrange{21}{100}{\percent}}
\newcommand{\InspiratoryPressureAlarmSedatedRange}{\SIrange{10}{80}{\cmw}}
\newcommand{\InspiratoryPressureAlarmSedatedStep}{\SI{1}{\cmw}}
\newcommand{\TidalVolumeSedatedRange}{\SIrange{50}{1500}{\milli\liter}} 
\newcommand{\TidalVolumeSedatedStep}{\SI{50}{\milli\liter}} 
\newcommand{\MinuteVentilationSedatedRange}{\SIrange{2}{20}{\slpm}} 
\newcommand{\MinuteVentilationSedatedStep}{\SI{1}{\slpm}}

\newcommand{\FractionInspiratoryFlowActiveRange}{\SIrange{5}{20}{\percent}}
\newcommand{\FractionInspiratoryFlowActiveStep}{\SI{1}{\percent}}

\newcommand{\MaxOxigenConsumption}{\SI{6}{\slpm}}

%% file: mvm.auth.tex
\author{C.~Galbiati}\affiliation{\PrincetonPHYAddress}\affiliation{\AQGSSIAddress}\affiliation{\AQLNGSAddress}\affiliation{\CentroFermiAddress}
\author{A.~Abba}\affiliation{\NuclInstruments}
\author{P.~Agnes}\affiliation{\UHoustonAddress}
\author{P.~Amaudruz}\affiliation{\TRIUMFAddress}
\author{M.~Arba}\affiliation{\CAINFNAddress}
\author{F.~Ardellier-Desages}\affiliation{\APCAddress}
\author{C.~Badia}\affiliation{\AQGSSIAddress}
\author{G.~Batignani,}\affiliation{\PIUniPHY}\affiliation{\PIINFNAddress}
\author{G.~Bellani}\affiliation{\MiBicoccaMEDAddress}
\author{G.~Bianchi}\affiliation{\STIIMACNRAddress}
% ...

%% file: mvm.inst.tex
\newcommand{\AQLNGSAddress}{INFN Laboratori Nazionali del Gran Sasso, Assergi (AQ) 67100, Italy}
\newcommand{\AQGSSIAddress}{Gran Sasso Science Institute, L'Aquila 67100, Italy}
\newcommand{\AZPneuAddress}{{AZ Pneumatica S.r.l., Misinto (MB) 20826, Italy}}
\newcommand{\APCAddress}{Universit\'e de Paris, CNRS, Astroparticule et Cosmologie, F-75013 Paris, France}
\newcommand{\BOINFNAddress}{INFN Sezione di Bologna, Bologna 40126, Italy}
\newcommand{\CAINFNAddress}{INFN Sezione di Cagliari, Cagliari 09042, Italy}
\newcommand{\CentroFermiAddress}{Museo della fisica e Centro studi e Ricerche Enrico Fermi, Roma 00184, Italy}
\newcommand{\FNALAddress}{Fermi National Accelerator Laboratory, Batavia, IL 60510, USA}
\newcommand{\GEINFNAddress}{INFN Genova, Genova 16146, Italy}
\newcommand{\UCLAAddress}{Physics and Astronomy Department, University of California, Los Angeles, CA 90095, USA}
\newcommand{\NAINFNAddress}{INFN Sezione di Napoli, Napoli 80126, Italy}
\newcommand{\NAUniPHYAddress}{Physics Department, Universit\`a degli Studi ``Federico II'' di Napoli, Napoli 80126, Italy}
\newcommand{\PrincetonPHYAddress}{Physics Department, Princeton University, Princeton, NJ 08544, USA}
\newcommand{\QueensPHYAddress}{Department of Physics, Engineering Physics and Astronomy, Queen's University, Kingston, ON K7L 3N6, Canada}
\newcommand{\RMUnoINFNAddress}{INFN Sezione di Roma, Roma 00185, Italy}
\newcommand{\RMTreINFNAddress}{INFN Sezione di Roma Tre, Roma 00146, Italy}
\newcommand{\RMUnoUniAddress}{Physics Department, Sapienza Universit\`a di Roma, Roma 00185, Italy}
\newcommand{\RMTreUniAddress}{Dipartimento di Matematica e Fisica, Universit\`a Roma Tre, Roma, Italy}
\newcommand{\STIIMACNRAddress}{CNR STIIMA, Milano 20133, Italy}
\newcommand{\MiBicoccaFis}{Physics Department, University of Milano - Bicocca, Milano 20126, Italy}
\newcommand{\UHoustonAddress}{Department of Physics, University of Houston, Houston, Texas 77204, USA}
\newcommand{\AstroCeNT}{AstroCeNT, Nicolaus Copernicus Astronomical Center, Polish Academy of Sciences, Warsaw 00-614, Poland}
\newcommand{\CNAF}{Istituto Nazionale di Fisica Nucleare - CNAF, Bologna (BO) 40127, Italy}
\newcommand{\UMassAddress}{Amherst Center for Fundamental Interactions and Physics Department, University of Massachusetts, Amherst, MA 01003, USA}
\newcommand{\ISTPAddress}{Istituto per la Scienza e Tecnologia dei Plasmi del CNR, ISTP-CNR, Milano 20125,  Italy}
\newcommand{\SanGerardoAddress}{Azienda Ospedaliera San Gerardo, Milano, Italy}
\newcommand{\SanPietroAddress}{Policlinico San Pietro, Ponte San Pietro, Italy}
\newcommand{\MiBicoccaMEDAddress}{Dipartimento di Medicina e Chirurgia, University of Milano - Bicocca, Milano 20126, Italy}
\newcommand{\MIPAddress}{Max-Planck-Institut f{\"u}r Physik (Werner-Heisenberg-Institut), F{\"u}ringer Ring 6, München 80805, Germany}
\newcommand{\DSIMAquilaAddress}{DISIM, Universit{\`a} de L'Aquila}
\newcommand{\LPNHEAddress}{LPNHE, CNRS/IN2P3, Sorbonne Universit\'e, Universit\'e Paris Diderot, Paris 75252, France}
\newcommand{\NuclInstruments}{Nuclear Instruments S.R.L., Como 22045, Italy}
\newcommand{\NAUniPHAAddress}{Department of Pharmacy, Universit\`a degli Studi ``Federico II'' di Napoli, Napoli 80126, Italy}
\newcommand{\MDSurgiMedicMolecUniPi}{Dep. of Surgical, Medical, Molecular Pathology and Critical Care Area,  University of Pisa, Pisa 56126, Italia}
\newcommand{\PIINFNAddress}{INFN Sezione di Pisa, Pisa 56127, Italy}
\newcommand{\twelfthmantec}{12th Man Technologies, Garden Grove 92841, US}
\newcommand{\CNLAddress}{Canadian Nuclear Laboratories,  Plant Rd Deep River, Canada}
\newcommand{\TRIUMFAddress}{TRIUMF - Canada's National Laboratory For Particle and Nuclear Physics , Vancouver V6T 1Z4, Canada}
\newcommand{\SNOLABAddress}{SNOLAB, Lively (ON) P3Y 1N2, Canada}
\newcommand{\DipPediatricsSapienzaAddress}{Department of Pediatrics, Sapienza Universit\`a di Roma, Roma 00185, Italy}
\newcommand{\MIStataleAddress}{Dipartimento di Fisica, Università di Milano, Milano 20133, Italy}
\newcommand{\MIINFNAddress}{INFN Sezione di Milano, Milano 20133, Italy}
\newcommand{\GinevriSRLAddress}{GINEVRI S.R.L., Via Cancelliera, 25/b, Albano Laziale 00041, Italy}
\newcommand{\ISSRoma}{National Center for Innovative Technologies in Public Health, ISS (Italian National Institute of Health), Roma 00161, Italy}
\newcommand{\EquilibarAddress}{Equilbar L.L.C., Fletcher 28732, USA}
\newcommand{\CIEMATAddress}{CIEMAT, Centro de Investigaciones Energ\'eticas, Medioambientales y Tecnol\'ogicas, Madrid 28040, Spain}
\newcommand{\PIUniPHY}{Physics Department, Universit\`a degli Studi di Pisa, Pisa 56127, Italy}
\newcommand{\CERNaddress}{CERN, European Organization for Nuclear Research 1211 Geneve 23, Switzerland, CERN}
\newcommand{\TorontoHospitalAddress}{St. Michael's Hospital - University of Toronto,Toronto M5B 1W8,CANADA}
\newcommand{\CAENAddress}{CAEN S.p.A., Viareggio 55049, Italy}
\newcommand{\SIUniPHY}{Dipartimento SFTA, Universit\`a degli Studi di Siena, Siena 53100, Italy}
\newcommand{\PoliMiDipMeccAddress}{Dipartimento di Meccanica, Politecnico di Milano, Milano 20156, Italy}
\newcommand{\ElemasterAddress}{Elemaster Group S.p.A., Lomagna (LC) 23871, Italy}
\newcommand{\OxfordAddress}{Department of Physics, University of Oxford, Oxford OX1 3RH, United Kingdom}
\newcommand{\MinesParisThecAddress}{Mines Paristech - PSL - Research University, Fontainebleau Cedex 77305, France}
\newcommand{\DipPhyRochesterAddress}{Department of Physics and Astronomy, University of Rochester, Rochester, NY 14627, USA}
\newcommand{\ParmeggianoAddress}{Sergio Installazioni Snc, Treviglio (BG) 24047, Italy}
\newcommand{\LSCAddress}{LSC - Laboratorio Subterráneo de Canfranc, Canfranc-Estación 22880, Spain}

%% file: sections/Introduction.tex
\section{Introduction}
\label{sec:Introduction}

The large number of people affected by \CovidPath\ has created an urgent demand for ventilators on a global basis, a demand that exceeds the capacity of the existing supply chains, especially in some regions where cross-border supply has been disrupted.  This need has motivated the development of the  mechanical ventilator (Mechanical Ventilator Milano, \MVM) -  a reliable, fail-safe, and easy to operate mechanical ventilator that can be produced quickly, at large scale, based on readily-available parts.  It was inspired by the Manley ventilator~\cite{Manley:1961do}, which was proposed in 1961, based on {\it ``the possibility of using the pressure of the gases from the anaesthetic machine as the motive power for a simple apparatus to ventilate the lungs of the patients in the operating theatre''}~\cite{Feldman:1995km}.  The \MVM\ is designed with the same principle of simplicity in mind.

The current version of this paper reflects up to date revisions stemming from  our testing and from recommendation by medical doctors.\\
However, we foresee possible future updates to this paper, as we proceed with testing and as we assess parts availability which may vary from country to country.
We are also proceeding with the required tests for certification of this ventilator, working with regulators in Italy and several other countries.

\MVM\ is designed to work in a pressure-controlled mode, which appears to be the correct operation mode for the COVID-19 patients, for whom a high pressure may damage further the lungs. \MVM\  and can be operated in both independent ventilation (pressure-controlled ventilation, \PCV) and patient-assisted control modes (pressure-supported ventilation, \PSV).

The system  connects directly to a line of pressurized medical oxygen or medical air, and relies on regulation of the flow to deliver medical air, medical oxygen, or a mixture of air and oxygen to the patient at a pressure in the range suitable for treatment.  Pressure regulation of the end-expiratory cycle is achieved by discharging the expiratory flow through a valve which sets the desired minimum positive end-expiratory pressure (\PEEP).  Another adjustable pressure limiting valve is connected to the inspiratory line and ensures that the maximum pressure delivered does not exceed the pre-set value.

Vocabulary and semantics are defined as per ISO 19223:2019~\cite{InternationalOrganizationforStandardization:2019wu}.  The system is designed to satisfy compliance with the guidelines defined in the international standard ISO 80601-2-12:2020~\cite{InternationalOrganizationforStandardization:2020vn}.  The most significant qualification tests in this respect are discussed in \refsec{TestISO}.

Important features of the \MVM\ are:

\begin{compactitem}
\item[\bf Small Number of Components:] as described in \refsec{Design}.
\item[\bf Ease of Procurement:] the parts required for the construction of the \MVM\ have been selected based on those that are available in many nations globally.  The parts selected are also characterized by their ease of use in large-scale manufacturing and assembly.
\item[\bf Simplicity of Construction:] assembly of the parts into a complete \MVM\ is achievable based on a small set of  clear instructions.  The process for loading the software into the controller is simple.  The controller software is open source and available for customization by end users.
\item[\bf Cost Containment:] the total cost of the components is in the hundred \EUR{}'s.
\item[\bf Convenience of Deployment:] the device requires only connection to a line of pressurized oxygen and standard AC electrical power (either \num{110} or \SI{220}{\V}); this makes the \MVM\ readily deployable in medical clinics with centralized oxygen and air supply systems, such as \Covid\ hospitals or \Covid-care areas in general hospitals.
\item[\bf  Customizability:] the \MVM\ can operate in different ventilation modes: independent (PCV) and patient-assisted (PSV) as described in \refsec{Controls}. Also, the operating parameters can be tuned by the operator with a simple user interface.
\item[\bf Reliability:]  the components in the \MVM\ are commercial and readily available.  We note that the reliability of \MVM\ has to be carefully studied based on the specifications of the components.  The system is designed to be easy to repair, just by replacing any non-functioning parts.
\item[\bf Limited oxygen consumption:] the consumption of oxygen with this device will not exceed \MaxOxigenConsumption.
\end{compactitem}

%% file: sections/Design.tex
\section{Design and Components}
\label{sec:Design}

The illustration in  ~\reffig{MVM-Illustration} shows the MVM ventilator (components within the light blue box) and a possible setup for the corresponding breathing circuit. The main components are described below:

\begin{figure*}[t!]
\centering{}
\includegraphics[width=\textwidth]{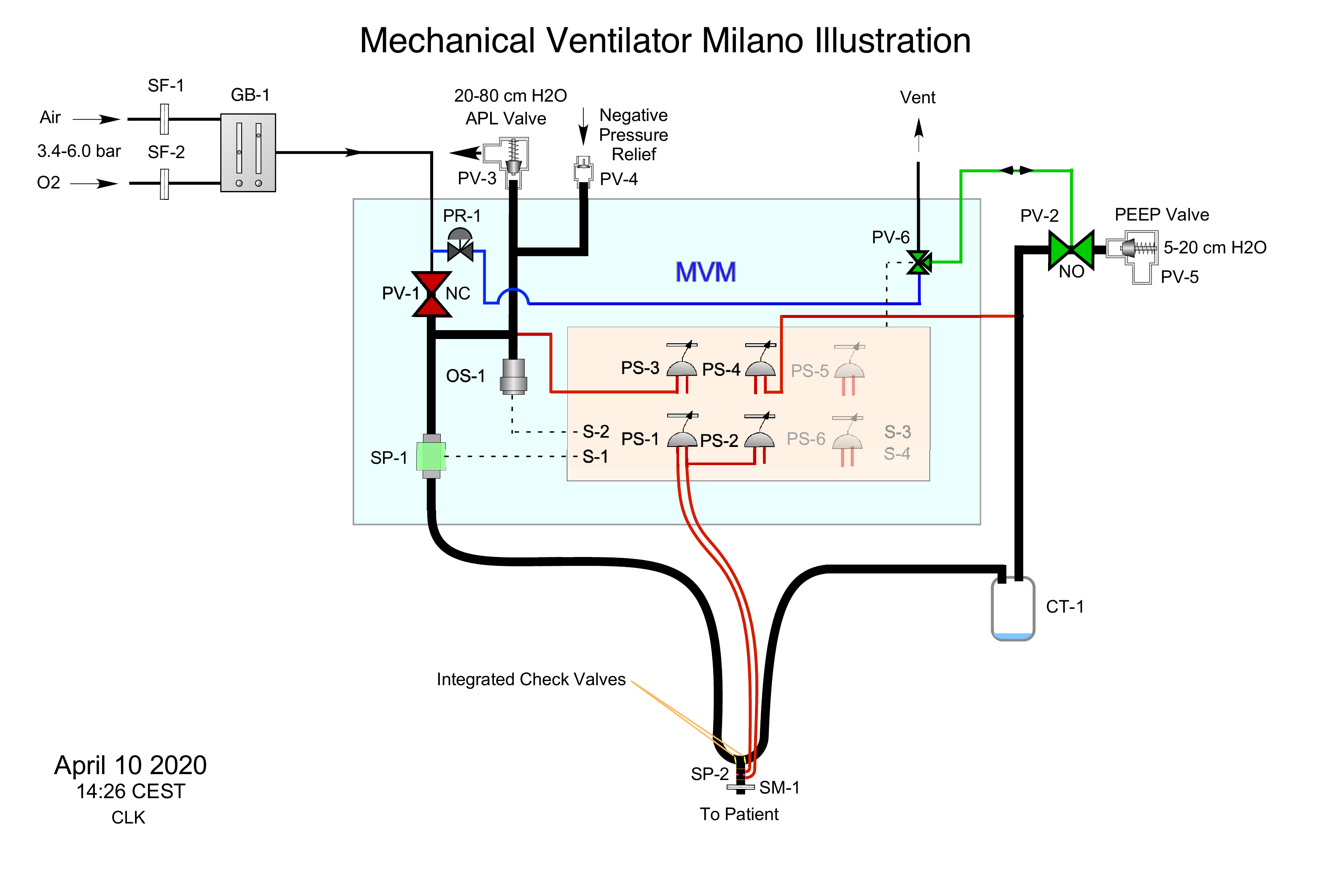}
\caption[\MVM\ Illustration.]{\label{fig:MVM-Illustration} Illustration of the \MVM\ ventilator and possible breathing circuit.}
\end{figure*}

\begin{figure*}[t!]
\centering 
\includegraphics[width=\textwidth]{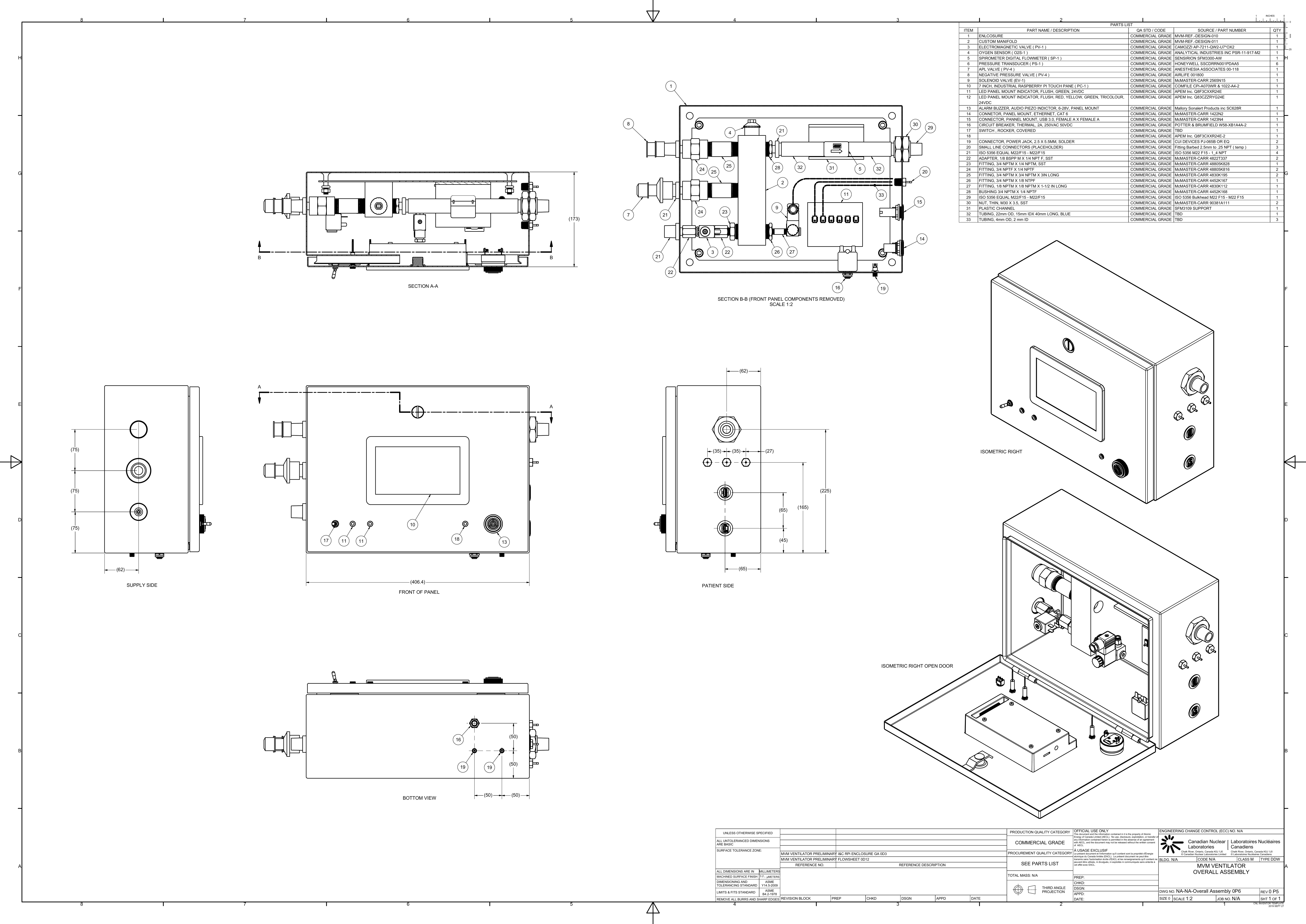}
\caption{\label{fig:MVM-Assembly} The \MVM\ controller base assembly.}
\end{figure*}

\begin{compactitem}
\item[\bf Connection to oxygen and air supply:]  at the left-hand side, the \MVM\ is connected to a pressurized oxygen/air line;
\item[\bf Sintered filters \SFOne, \SFTwo] the sintered filters remove particulate in the inline that can clog the pipes;

\item[\bf Gas blender \GBOne] The oxygen and air flow are mixed in a medical gas blender.  \GBOne\ is external  to the \MVM\ unit.  The \FiOTwo\ set point 
is set 
manually, directly on the \GBOne\ unit.

\item[\bf Differential pressure sensors:] Four differential pressure sensors PS-1, PS-2, PS-3, and PS-4 monitor the pressure and flow at different points of the breathing system.  
\item[\bf Air/Oxygen delivery proportional valve \PVOne:] The incoming gas flow to the patient is controlled by the proportional valve \PVOne\ using a process control loop based on the value of PS-1 to ensure that the proper respiratory minute volume is delivered to the patient.  \PVOne\ is a normally close (\NC) valve;
\item[\bf Adjustable pressure limiting valve \PVThree:]  Mechanical valve that sets the value of the maximum inspiratory pressure in the range \APLPressureRange;
\item[\bf Negative pressure relief valve \PVFour:] A check valve to avoid any negative pressures during patient assisted ventilation. In that mode, the patient is active  and can spontaneously request more air. As long as there is a positive pressure in the respiratory lines this valve  remains closed. A bacterial/viral filter  may be required just before the valve to prevent contaminated room air entering the upstream part of the system;
\item[\bf Oxygen sensor \OSOne:] An oxygen sensor \OSOne\ is used to continuously monitor the fraction of inspired oxygen \FiOTwo;
\item[\bf Spirometer \SPOne:] A precision spirometer is connected to the input line to monitor the inspiratory flow rate;
\item[\bf Breathing system:] The breathing system, connected to the tracheal tube, supports the attachment of two plastic tubes of standard size \MaskTubeSize\ connecting respectively to valves \PVOne\ and \PVTwo, and to a smaller plastic tube leading to the differential pressure sensors \PSOne, \PSTwo, \PSThree, and \PSFour\ to monitor pressure and flow.  The standard for connection of the breathing system is the \MaskTubeSize\ cone and socket combination defined in the standard~\cite{InternationalOrganizationforStandardization:2015vh}; 
\item[\bf Condensate trap \CTOne:] The expiration tube passes through a condensate trap allowing for removal of condensed vapor from the patient's breath;
\item[\bf Silicone membrane \SMOne:] The silicone membrane filters access to the machine of the wet flow;
\item[\bf Expiration  valve \PVTwo:] This high throughput valve with low pressure-drop controls the expiratory flow.  An adequate orifice diameter guarantees the flow corresponding to the expiration of a proper respiratory minute volume at the given \PEEP\ values; the valve is controlled by the three-way solenoid valve \PVSix.
\item[\bf PEEP valve \PVFive:] A mechanical valve that controls and defines the positive end-expiratory pressure \PEEP\ in the range \PEEPPressureRange.
\end{compactitem}
A first technical layout  of the \MVM\ controller base assembly box is shown in \reffig{MVM-Assembly}.

%% file: sections/Controls.tex
\section{Control System and Operation}
\label{sec:Controls}

The control system performs supervision and actuation of the two valves (\PVOne\ and \PVTwo) based on the programmed respiratory cycle.

The pressure sensor \PSOne 
measures the pressure at the patient. The control system reads the pressure on \PSOne  and  appropriately adjusts the two valves. This ensures that the pressure is always within the operating range: \APLPressureRange\ for the inspiratory phase and \PEEPPressure\ for the expiratory phase.

During the inspiratory phase, \PVTwo\ is closed and \PVOne\ is open.  The setting of \PVOne\ is adjusted to maintain  the pressure  within the desired range during the inspiratory period.  At the end of the programmed inspiratory period, \PVOne\ is closed and, after a small pause, \PVTwo\ is opened to allow the discharge of the lung pressure.  The expiratory pressure limit PEEP is set by \PVFive.

The main controller runs on a ({\tt Arduino}-compatible) micro-controller board based on a \SI{32}{\bit} micro-controller.  These boards have a small form factor and integrate all I/O functions required for this system.  

A daughter board interfaces to the controller and provides four opto-coupled switches to operate the electrically-controlled valves. The daughter board is provided with \SI{24}{\volt} VDC supply and includes the low voltage regulator to supply the central unit. 

The differential pressure gauges are based on a \PSOneDiffPressure\ differential pressure sensor with a resolution better than \PSOneResolution.  

A buzzer and a high luminosity LED are incorporated to signal alarms.

The \MVM\ is equipped with an industrial power supply unit capable of at least \SupplyPower\, and battery backup operated in fail-safe mode.  Under normal circumstances, the power supply will feed the controller and keep the battery charged. In case of power failure, the battery will automatically provide the power for ongoing system operation for up to \BatteryLifetime.  The power supply is hosted in a separate enclosure to provide isolation between the oxygen lines and possible spark sources.  The backup power supply unit will include two \BatteryVoltage, \BatteryCharge\ batteries.

\subsection{Pressure-Controlled Ventilation Mode}
\label{sec:PCV}

In the pressure-controlled ventilation (\PCV) configuration, the unit will operate the valves in regular cycles. The operator defines the cycle by setting the inspiratory time, \PEEP, and respiratory rate.  The operator is also required to set the target inspiratory pressure. The maximum inspiratory pressure threshold is manually set by \PVThree\ for patient safety. Alarms are set on the basis of the inspiratory pressure, the minute ventilation, and the tidal volume, measured by the system itself.  Preliminary ranges of values for control and alarm parameters are listed in \reftab{PCV-Parameters}. 

\begin{table*}[t!]
\begin{center}
\begin{tabular}{lcc}
\hline\hline
\bf Control Parameter           &\bf Range                              &\bf Control Step\\
\hline
Respiratory rate                &\RespiratoryRateSedatedRange           &$\pm$\RespiratoryRateSedatedStep\\ 
Inspiratory time                &\InspiratoryTimeSedatedRange           &$\pm$\InspiratoryTimeSedatedStep\\  
\PEEP                           &\PEEPSedatedRange                      &\EMC\\    
Max inspiratory pressure        &\InspiratoryPressureMaxSedatedRange    &\EMC\\    
\FiOTwo                         &\FiOTwoSedatedRange                    &\EMC\\    
\hline
\bf Alarm Parameter             &\bf Range                              &\bf Control Step\\
\hline
Inspiratory pressure            &\InspiratoryPressureAlarmSedatedRange  &$\pm$\InspiratoryPressureAlarmSedatedStep\\
Tidal volume                    &\TidalVolumeSedatedRange               &$\pm$\TidalVolumeSedatedStep\\
Minute ventilation              &\MinuteVentilationSedatedRange         &$\pm$\MinuteVentilationSedatedStep\\
\hline
\end{tabular}
\caption[Control and alarm parameters for the \PCV\ mode of operation.]{Ranges of values for control and alarm parameters for the \PCV\ mode of operation of the \MVM.  (Note: \EMC\ stands for external mechanical control, typically achieved via spring-loaded \PEEP\ valves.)}
\label{tab:PCV-Parameters}
\end{center}
\end{table*}

\subsection{Pressure-Supported Ventilation Mode}
\label{sec:PSV}

In pressure-supported ventilation (\PSV) mode, the patient triggers the ventilator. Inspiration pressure support is given at a preset constant pressure. The ventilator regulates the pressure during inspiration so that it corresponds to preset values within the operating ranges listed in \reftab{PSV-Parameters}.

\begin{table*}[ht]
\begin{center}
\begin{tabular}{lcc}
\hline\hline
\bf Control Parameter           &\bf Range                              &\bf Control Step\\
\hline
\PEEP                           &\PEEPSedatedRange                      &\EMC\\    
Max inspiratory pressure        &\InspiratoryPressureMaxSedatedRange    &\EMC\\    
\FiOTwo                         &\FiOTwoSedatedRange                    &\EMC\\    
Fraction of inspiratory flow    &\FractionInspiratoryFlowActiveRange    &$\pm$\FractionInspiratoryFlowActiveStep\\
\hline
\bf Alarm Parameter             &\bf Range                              &\bf Control Step\\
\hline
Inspiratory pressure            &\InspiratoryPressureAlarmSedatedRange  &$\pm$\InspiratoryPressureAlarmSedatedStep\\
Tidal volume                    &\TidalVolumeSedatedRange               &$\pm$\TidalVolumeSedatedStep\\
Minute ventilation              &\MinuteVentilationSedatedRange         &$\pm$\MinuteVentilationSedatedStep\\
\hline
\end{tabular}
\caption[Control and alarm parameters for the \PSV\ mode of operation.]{Ranges of values for control and alarm parameters for the \PSV\ mode of operation of the \MVM.  (Note: \EMC\ stands for external mechanical control, typically achieved via spring-loaded \PEEP\ valves.) }
\label{tab:PSV-Parameters}
\end{center}
\end{table*}

%% file: sections/Electronics.tex
\section{Electronics}
\label{sec:readout}

The goals of the \MVM\ electronics are:

\begin{compactitem}
\item Control the valve system; 
\item Read the pressure and the oxygen sensors;
\item Generate hardware (LED, Buzzer) and visual alarms;
\item Provide a simple Graphic User Interface (GUI). 
\item Visualize the system parameters on a display. 
\end{compactitem}

\begin{figure*}[t!]
\centering{}
\includegraphics[width=\textwidth]{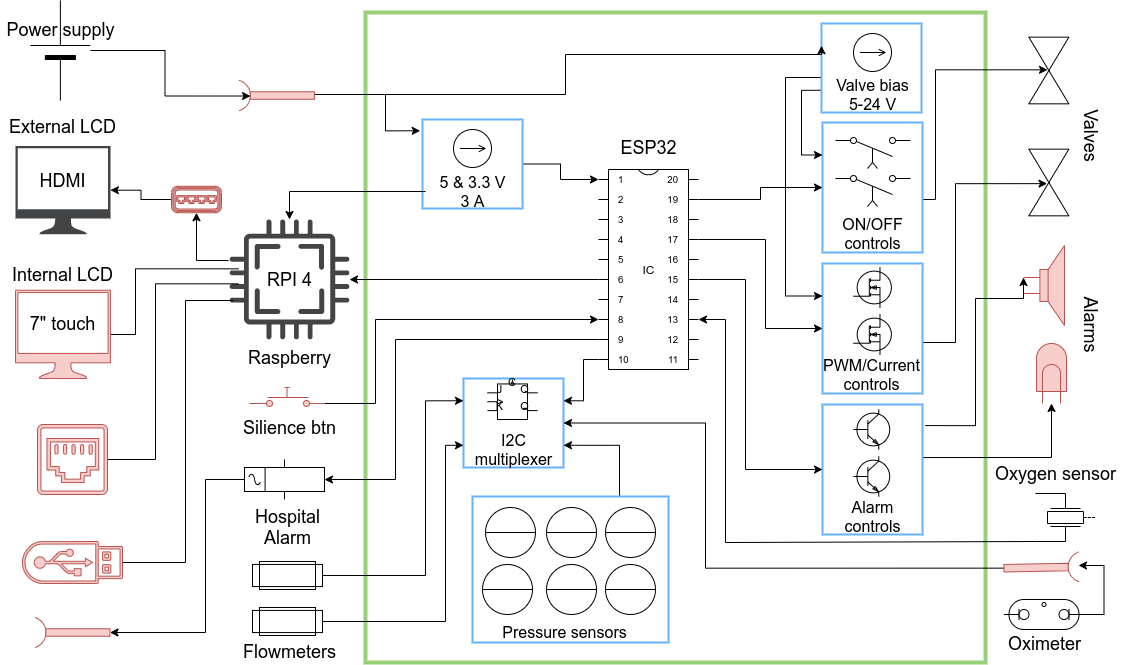}
\caption[]{\label{fig:MVM-Controls} Block diagram of the electrical connections: the green box defines the custom control board.}
\end{figure*}

It is composed by a board hosting an ESP32 micro-controller and a Raspberry~4 unit.  The parameters are displayed on a 7” touch-screen that allows also a few parameter settings. \\
\reffig{MVM-Controls} shows the block diagram of the electrical connections.

\subsection{The electronic board}

The \MVM\ operations are managed by an electronic system which includes all the components to measure relevant quantities, to drive the solenoid and proportional valves and to interface to the user via a touchscreen. The micro-controller unit is based on a commercial product by Adafruit within the Feather line: these units all share the same form factor and connections pinout while offering different micro-controller and connectivity options.  Moreover a set of daughter boards, called Wings, provide extensions (Ethernet, touchscreen displays, LoRa radios, etc.): the wings can be stacked.  For the MVM project two feathers are of interest:

\begin{compactitem}
\item The Huzzah32 unit is based on a 32 bit micro-controller (ESP32) produced by Espressif. The ESP32 includes a dual core 240 MHz LX6 micro-controller, 0.5 MB of RAM, Wi-Fi and BT connectivity. The Huzzah32 provides 21 GPIO 3.3 V (2 I2C, 1 SPI and 1 UART busses, 2 true DACs and 12 ADCs). Furthermore all I/O pins can be configured for PWM;
\item The M4 Express unit is based on a 32 bit micro-controller (ATSAMD51) produced by Microchip. The ATSAMD51 is based on a Cortex M4 core running at 120 MHz with 0.2 MB of RAM, 21 GPIO 3.3 V (up to 6 between I2C, SPI and UART, plus 2 true DACs, and 12 ADCs).
\end{compactitem}

Both Feathers can be programmed with Arduino. Between the two units we picked the ESP32-based solution given the more widespread use of this micro-controller in the IOT environment. This in turns corresponds to a larger availability of the units.

The control board has been developed at LNGS, according to the design requirements of the Milano team that is testing the proposed solution in the field. The control boards include the following sub-systems:

\begin{compactitem}
\item The connection to the Feather micro-controller and the voltage regulators. From the main power supply 5 and 3.3 V are obtained via switching buck regulators providing up to 3 A. The 5 V is then forwarded to the raspberry via a USB-C connector. The Feather connectors provide access to all 21 GPIO pins;
\item Up to 6 differential pressure sensors, model 5525DSO-DB001DS. The sensors are connected via a I2C multiplexer;
\item Two voltage regulators for the valves: since the operation voltage of the solenoid changed few times during the project, a proper regulation was required. Therefore a 3 A step-down buck regulator can provide any voltage below in the range 3-11 V, while a step-up boost regulator can provide up to 24 V 3 A;
\item Two ON/OFF opto-coupled valve control;
\item Two analog controlled valves. The circuit allows both PWM modulation and current control:
\item Two button input and high-power two alarm outputs (for buzzer and LED).
\item Four I2C ports (2 of which can operated with 5 V equipment) all connected via a multiplexer, 1 SPI port and 1 UART port, 3 analog ports. 
\end{compactitem}

Up to two I2C are reserved to the gas flow meters (Sensirion SFM3019), and other is reserved to the oxymeter. The oxygen sensor has not yet been identified yet: however several readout options are available.

The control firmware runs on the micro-controller: the firmware implements a state machine whose transitions are defined based on the sensor inputs and on the commands received from the GUI (via the Raspberry unit). The state machine in turn controls the phase of the respiratory cycle (inhalation, pause, exhalation) by operating on the valves. The \PVOne\ valve is controlled via a double feedback system: the first PID loop ensures that the \PSThree\ sensor always matches the set value (Pin). The second feedback programs the Pin value to maintain the breathing cycle at the required pressure measured at the mask level (sensor \PSTwo). The firmware also continuously monitors the gas flow meter (\PSOne) to decide when to interrupt the inspiration phase. Both  manual and the assisted operation modes are available.

The Raspberry 4 micro-computer is connected to the micro-controller unit via a micro-usb cable: this allow to reprogram the micro-controller easily via the serial line. The Raspberry unit communicates with the micro-controller with the serial line during normal operations, accessing the sensor parameters and configuring the set-points. A 7” touchscreen is connected to the Raspberry: a proper GUI is under development to interact with the user.

Alarms are issued directly by the ESP32 firmware or by the GUI.  The microcontroller unit  monitors the behavior of the connected sensors and will generate alarms when a condition happens such that the normal operation cannot be maintained. The possible alarms include:

\begin{compactitem}
\item Sensor not working properly
\item Pressure level cannot be reached or goes in overflow 
\item Exhaust line reaches differential pressure equal to zero (PEEP valve not 
  working.)
\item power cut: the system is running in battery mode
\end{compactitem}

The above alarms will be notified by a high luminosity LED and by a buzzer.  A push button is available to silence the alarm. The corresponding alarm code will be displayed in the LCD screen.
Furthermore the GUI will monitor the following parameters and will generate alarms in case the parameters will exceed the selected boundaries:

\begin{compactitem}
\item {\bf Inspiratory pressure}: the pressure applied during the inspiration phase to the patient
\item {\bf Vtidal (tidal volume)}: Volume of gas provided to the patient in a respiration cycle
\item {\bf MVe (respiratory minute volume)}:  Volume of gas inhaled (inhaled minute volume) or exhaled (exhaled minute volume) from a person's lungs per minute
\item {\bf \FiOTwo\ (fraction of inspired oxygen)}: Concentration of oxygen in the gas mixture that the patient inhales
\end{compactitem}

\subsection{The Graphic User Interface (GUI)}

\begin{figure*}[t!]
\centering{}
\includegraphics[width=0.49\textwidth]{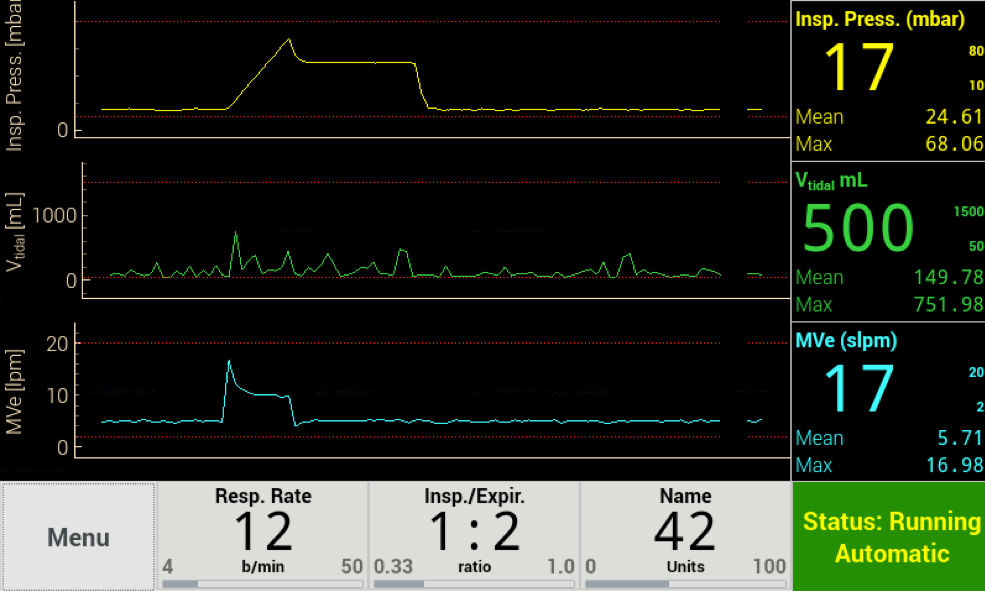}
\includegraphics[width=0.49\textwidth]{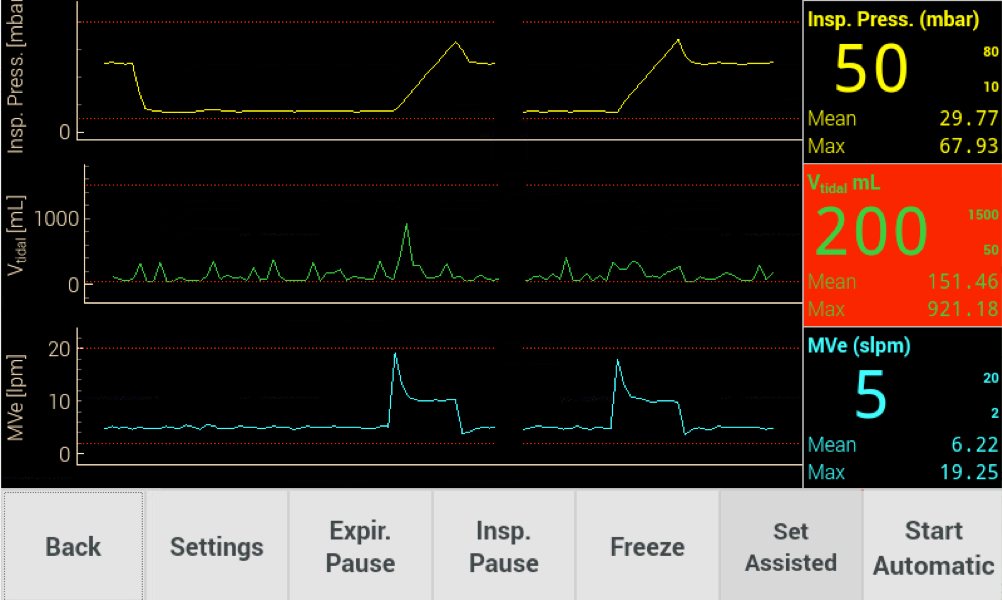}
\caption[]{Home screen, closed menu, no alarms(left) and V$_{tidal}$ (right) alarm. Preliminary version. Reported values may be unrealistic}
\label{fig:MVM-GUI}
\end{figure*}

The \MVM\ \GUI\ is a Python3 software, written using the PyQt5 toolkit, that allows steering and monitoring the \MVM\ equipment.
 \reffig{MVM-GUI} shows two screenshots of it.
 
Project design principles:
\begin{compactitem}
\item Ease of use and interface simplicity to make it immediate to understand and to give a familiar feeling by the user;
\item Use of entirely open-source software to let the project be easily spreadable and adaptable to different possible needs of users in other countries and to different approach to medical procedures;
\item Agile development techniques to speed up all phases of development (design, code, documentation, testing) and allow contributions from people having different skills;
\item Portability by default to support possible hardware platform and software environment (Operating System, OS) changes with minimal effort;
\item Support distributed input devices such as touchscreen, mouse and keyboard.
\end{compactitem}

Involved technologies:

\begin{compactitem}
\item Target computing platform: Raspberry Pi 4 (any memory size), chosen as a trade-off between its computing power over power consumption ratio and its wide availability on the market;
\item Target operating: Raspbian version 2020-02-13;
\item Target programming language: Python 3.5;
\item Target PyQt5: version 5.11.3.
\end{compactitem}

The \MVM\ \GUI\ runs smoothly on the target hardware and software environment.  

%% file: sections/TestComp.tex
\section{Comparison of \MVM\ with Siemens Servo900}
\label{sec:TestComp}

Tests were  performed with a breathing simulator, \ALSFiveK\ of \IngMar~\cite{IngMarMedical:2017vt}. \\
\begin{figure*}[ht!]
\centering{}
\includegraphics[width=0.49\textwidth]{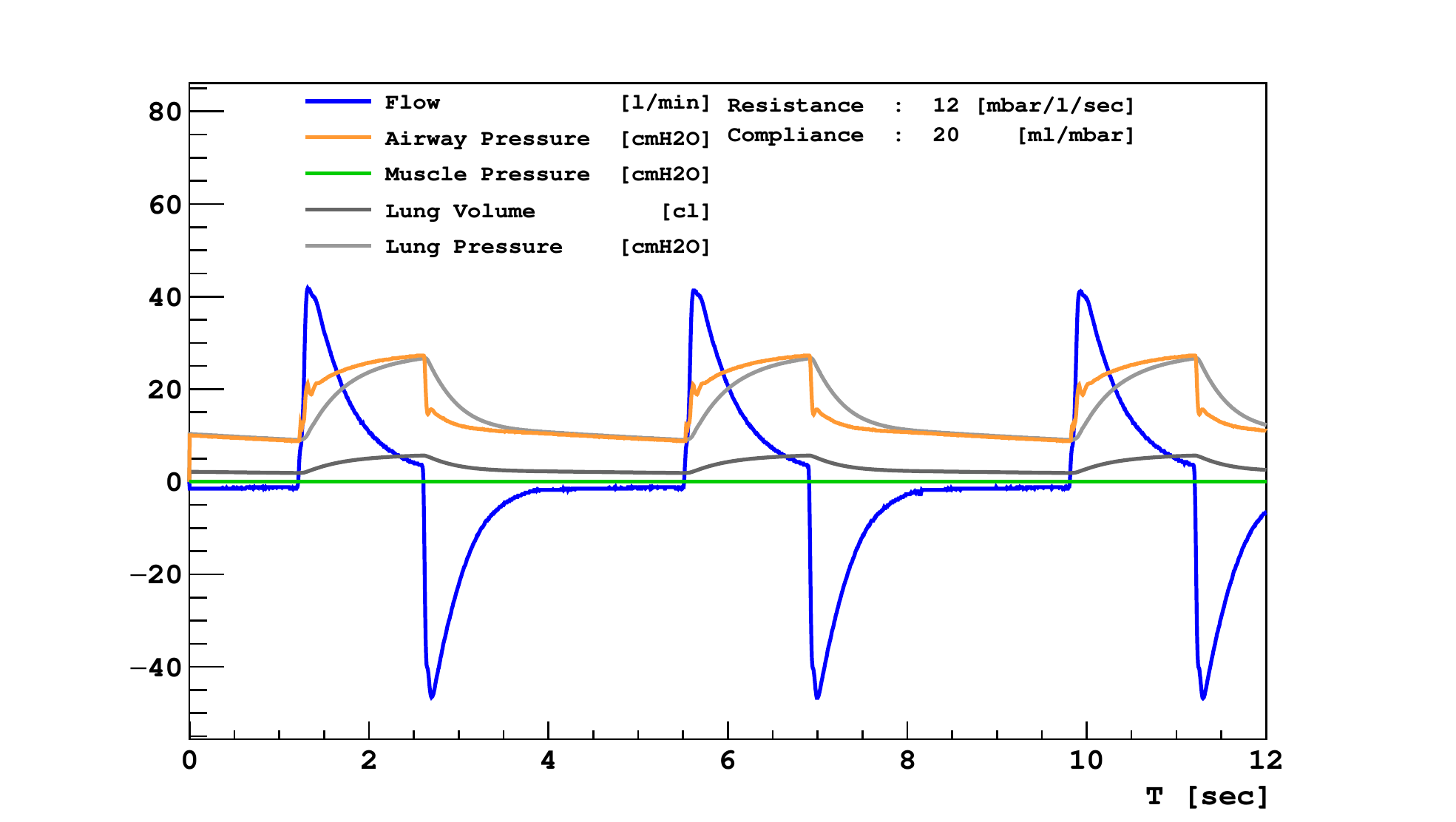}
\includegraphics[width=0.49\textwidth]{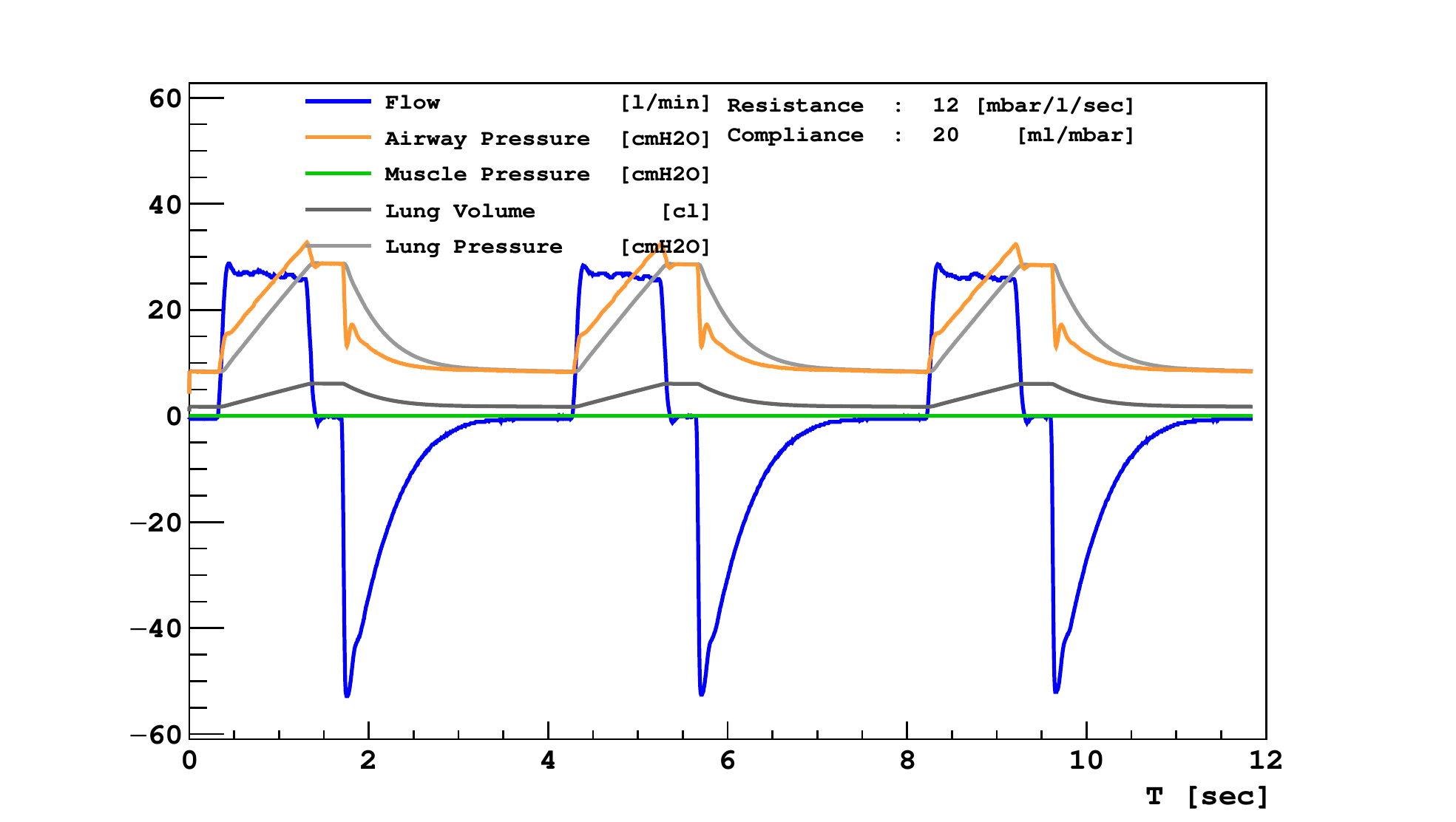}
\includegraphics[width=0.49\textwidth]{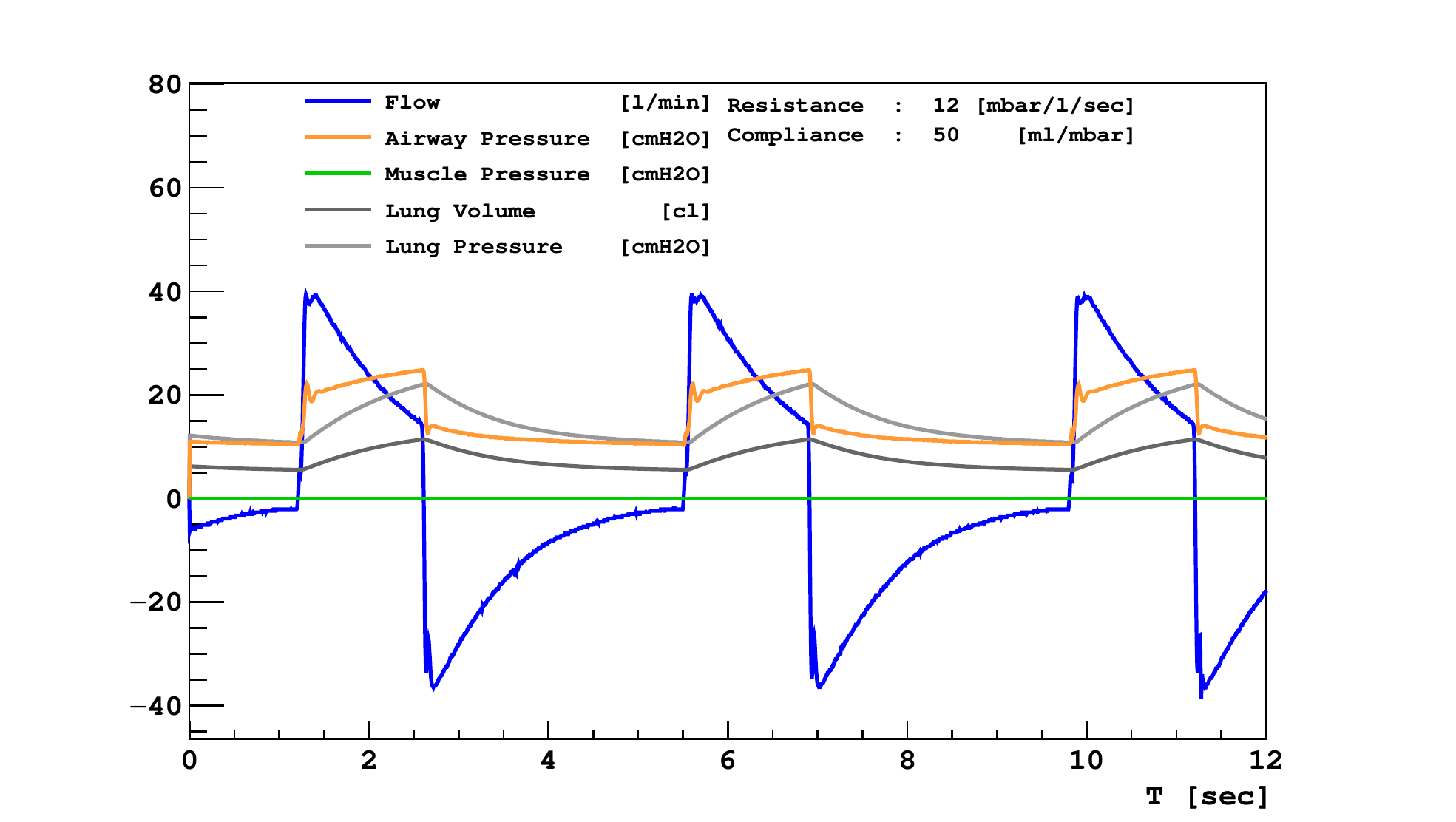}
\includegraphics[width=0.49\textwidth]{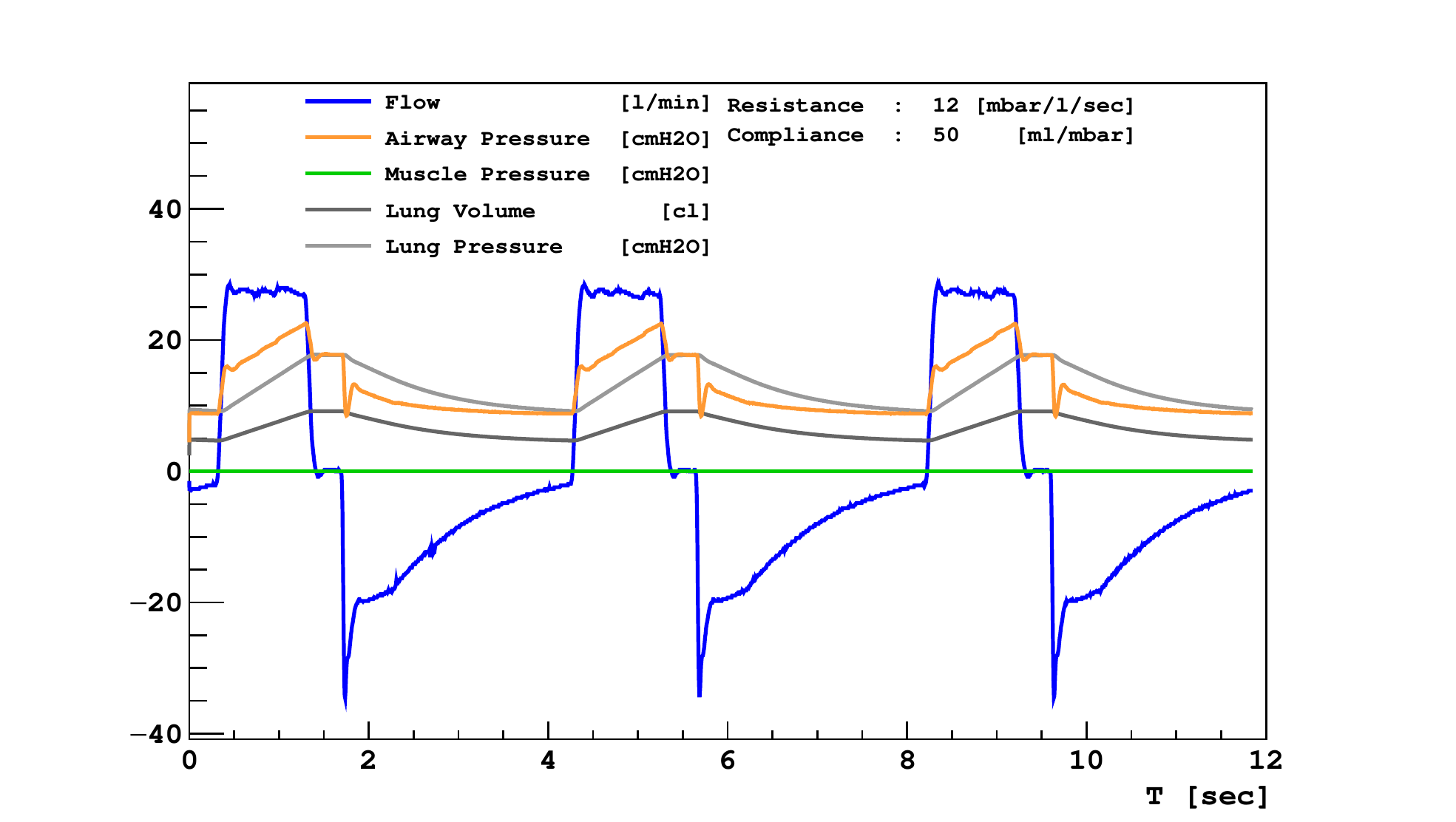}
\includegraphics[width=0.49\textwidth]{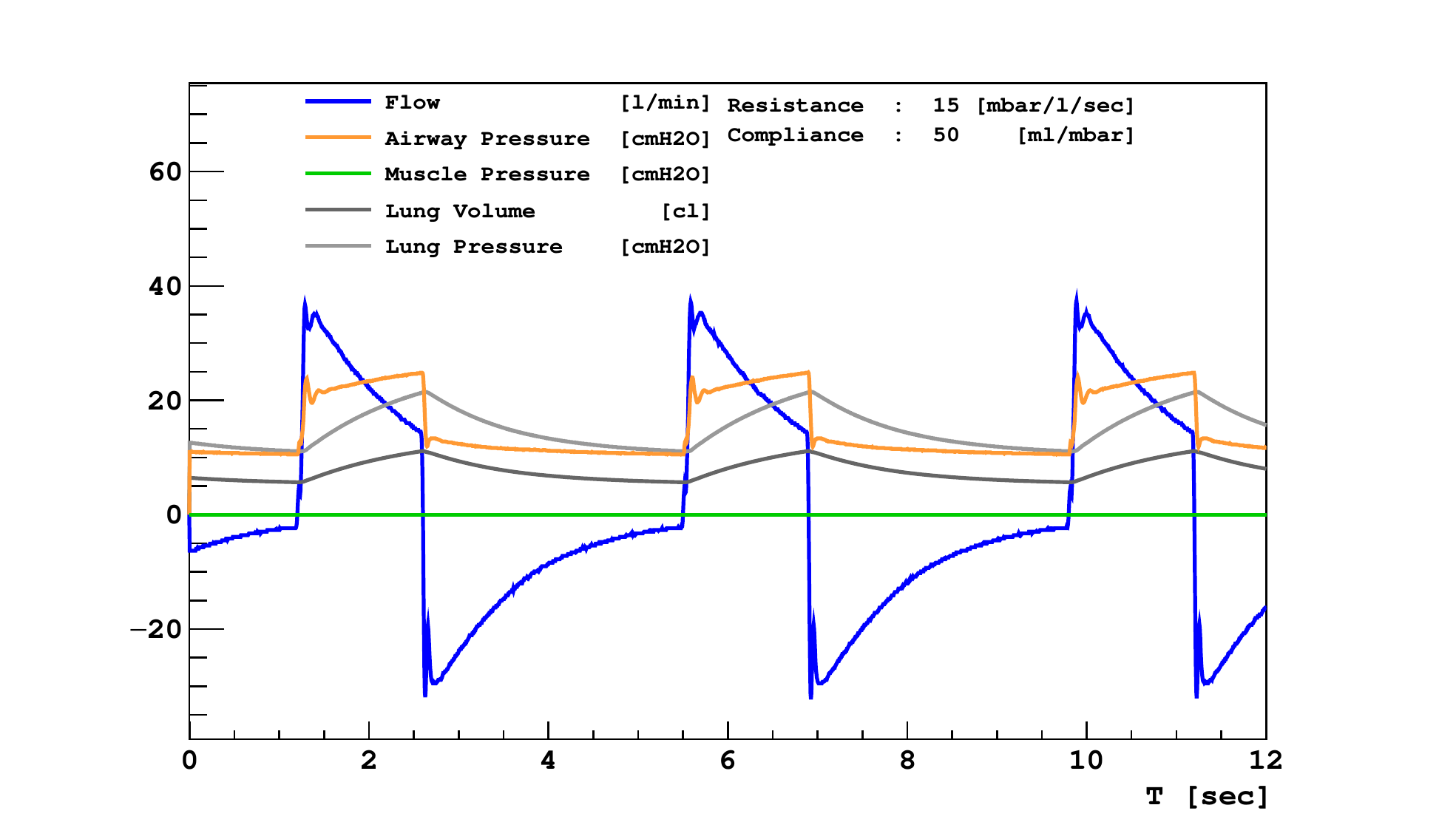}
\includegraphics[width=0.49\textwidth]{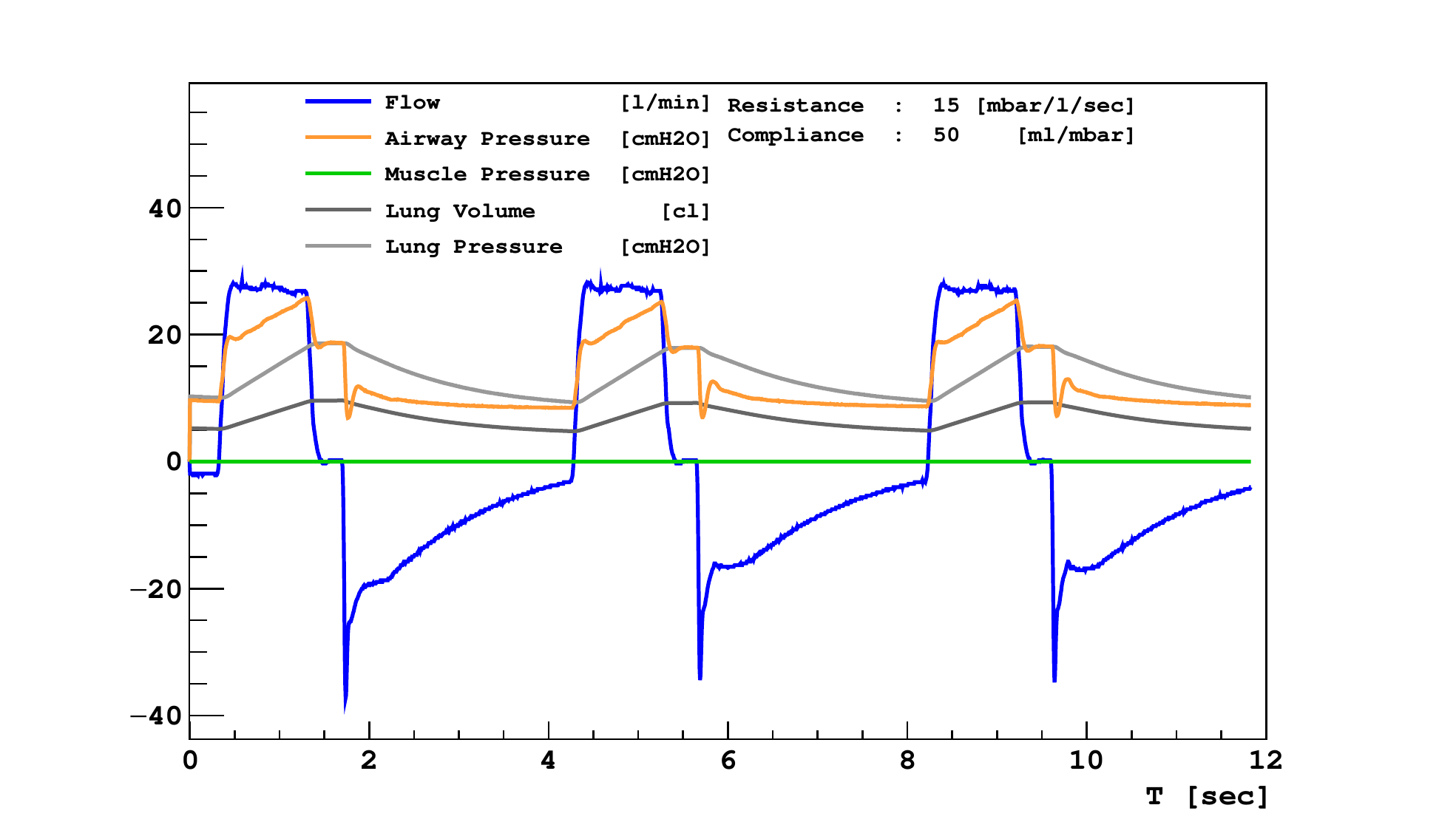}
\caption[Comparison of performance of the \MVM\ and of a Siemens Servo900 ventilator.]{Comparison of performance of the \MVM\ (left column) and of a Siemens Servo900 ventilator (right column).  See text for details.}
\label{fig:TestComp-MVM-Servo} 
\end{figure*}

\reffig{TestComp-MVM-Servo} shows the waveform comparison in the \PCV\ mode of \MVM\ performance compared to that of a \Siemens\ \ServoNineC\ device, a commercial ventilator working in flux control mode.  The \MVM\ operates in pressure control mode.  The difference in operation mode of the two ventilators is clearly visible in the shape of the flow and of the airway and lung pressure. 

In the \MVM\ the airway pressure curve take some time to reach the saturation value due to impedance of the tube connecting the pressure sensors of the \MVM\ to the simulator.  The \MVM\ precision spirometer is unidirectional and is therefore able only to measure the inspiratory flow.  The measurement of the expiratory flow is measured with a lesser precision by the Venturi spirometer \SPTwo\ and not shown in \reffig{TestComp-MVM-Servo}.

As an example, let's consider the top two plots of \reffig{TestComp-MVM-Servo}. On the right-hand side, the \ServoNineC\ ventilator maintains a steady, but convoluted with a clear oscillatory pattern, flow. Both airway and lung pressures experience a linear increase in time. 

On the left-hand side, the \MVM, operating in a pressure-controlled ventilation (PCV) mode, aims at maintaining a constant pressure, but while simultaneously implementing two crucial precautions. First, the flow is maximised at the beginning of the inspiratory cycle: this is the most appropriate procedure for \Covid\ patients as it allows the immediate reopening of the alveoli, and is strongly recommended by the doctors and nurses in the \Covid\ wards of Lombardy, rather than the constant flow procedure. Second, following the immediate increase in airway and lung pressure in coincidence with the start of the inspiratory cycle, the pressure is raised gently for the last fifteenth percentile of the full pressure differential (pressure differential is the difference between set inspiratory pressure and \PEEP) through the end of the inspiratory cycle. The reader will also notice that, during the inspiratory cycle, through the last  fifteenth percentile of the full pressure differential, the pressure rises smoothly and without any sort of oscillatory pattern through. 

These superior characteristics of the \MVM\ pressure transient during the inspiratory cycle are crucial to avoid barotrauma and to minimise long term fatigue of muscles and alveoli induced by forced mechanical ventilation.  This special \MVM\ pressure transient is achieved thanks to its ability to finely tune the airway and lung pressure by controlling the input pressure through a complex yet fail-safe feedback control loop. 

In essence, by listening to the precious advice of medical doctors and nurses in the \Covid\ wards of Lombardy, we implemented in the \MVM\ what we believe to be the safest and most beneficial pressure transient for treatment of \Covid\ patients.

%% file: sections/TestISO.tex
\section{Tests based on the ISO 80601-2-12:2020 testing protocol}
\label{sec:TestISO}

This section presents the results of the test required by the ISO reference standard~\cite{InternationalOrganizationforStandardization:2020vn}, section 201.12 for Pressure Controlled inflation-type testing, subsection 201.12.1.102.

Our test setup is equivalent to that described in figure 201.102 of the same documents.  The measurement performed with the \MVM\ are those reported in table 201.105.  We performed the measurements over 30 cycles in accordance to the prescription of the reference standard~\cite{InternationalOrganizationforStandardization:2020vn}

The \MVM\ unit does not include an internal oxygen control unit.  Instead, the \FiOTwo\ level can be independently set by using gas blender \GBOne, external to the unit, controlled either manually or directly by the \MVM\ unit.  For this reason, the \FiOTwo\ level is not varied during the tests as control of its value pertains to the performance of \GBOne\ as opposed to the performance of the \MVM.  Therefore, for this round of measurements the output value of the oxygen sensors \OSOne\ is not reported as the tests are done with air and the measured concentration of oxygen is constant at \SI{21}{\percent}.


Tidal volume is displayed both measured by the simulator and for the \MVM.  The \MVM\ measures the tidal volume by integration of the inspiratory and expiratory flows.  For simplicity of rendering of the data, the direct measurement of the expiratory flow is not shown.  To best support the understanding of the performance of the \MVM, data collected with the \MVM's own sensors are shown along with the data collected with the \ALSFiveK\ breathing simulator.  Pressure and \PEEP\ are measured according to the requests in the last \SI{50}{\ms} respectively of the inspiratory and expiratory phase.  A humidifier was not included in this test as its presence is not expected to affect qualification of the \MVM\ performance.

The recorded waveforms are presented in \reffig{ISOTest-1} to \ref{fig:ISOTest-4}.

\input{latex_test_ISO}

\clearpage 

\subsection{Summary plots}

\begin{figure}[ht!]
 \centering
    \begin{subfigure}[t]{0.49\linewidth}
        \includegraphics[width=\linewidth]{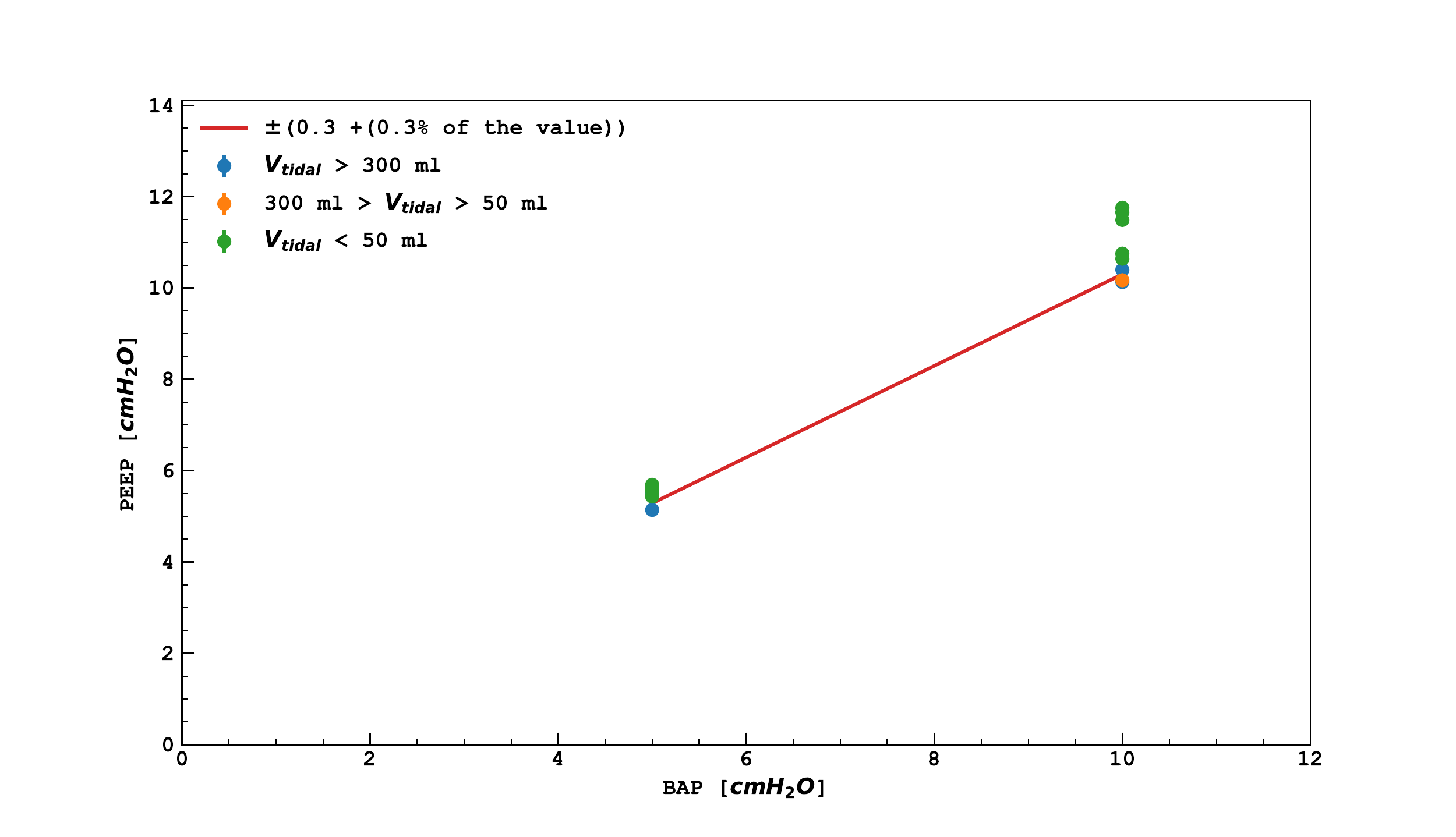}
        \caption{Measured PEEP in the last 50ms of the expiratory phase vs. BAP}
    \end{subfigure}
    \begin{subfigure}[t]{0.49\linewidth}
        \includegraphics[width=\linewidth]{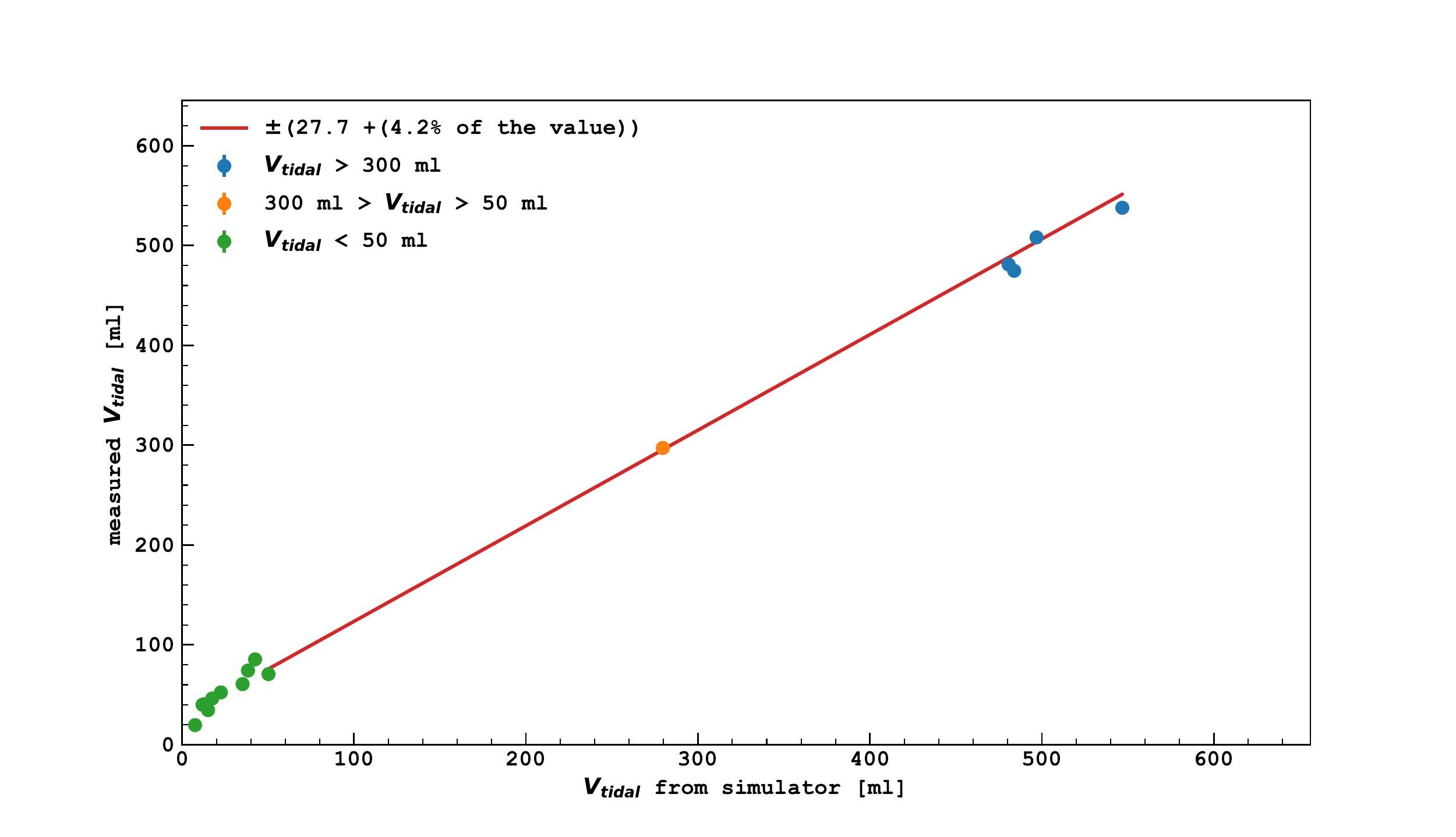}
        \caption{Measured tidal volume (V$_{tidal}$) by MVM vs simulator.}
    \end{subfigure}
    \begin{subfigure}[t]{0.49\linewidth}
        \includegraphics[width=\linewidth]{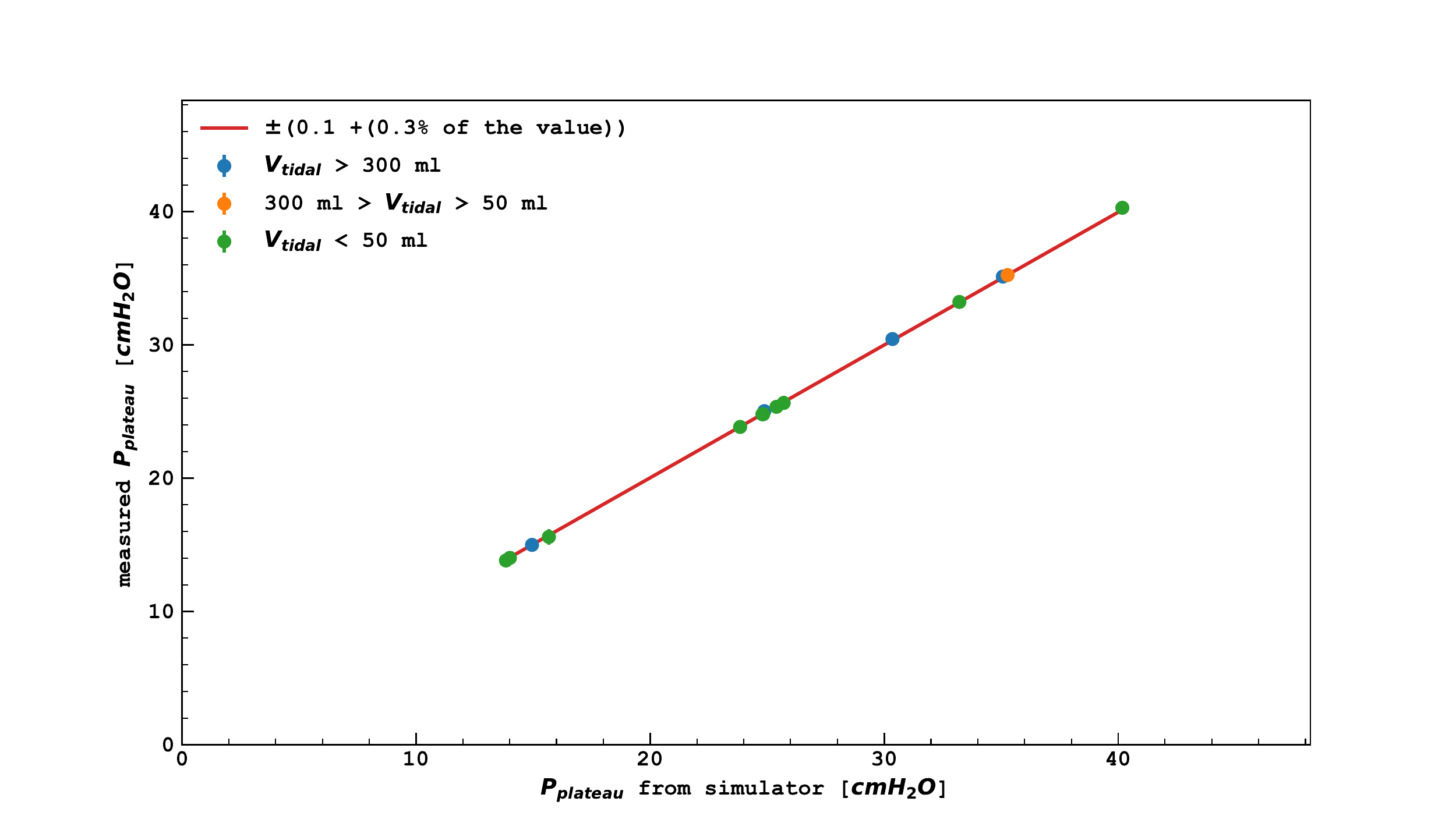}
        \caption{Measured pressure in the last 50ms at the plateau of the inspiratory phase (P$_{plateau}$) by MVM vs simulator.}
    \end{subfigure}
  
    \caption{Measured quantities as extracted from the measured waveforms vs set quantities. The average values and the maximum errors are calculated by separating them in groups of tidal volumes (V$_{tidal}$) as requested by the ISO document. }
      \label{fig:summary}
\end{figure}

Measured quantities as extracted from the measured waveforms vs set quantities are reported in \reffig{summary}. The average values and the maximum errors are calculated by separating them in groups of tidal volumes (V$_{tidal}$) as requested in~\cite{InternationalOrganizationforStandardization:2020vn}.

Linear fits were performed to extract the accuracies on the various parameters.
The results show that 
\begin{itemize}
    \item the measured P$_{plateau}$ accuracy is  $\pm$(0.3 +(0.3\% of the value))\si{\cmw}
    \item the measured tidal volume accuracy is $\pm$(27.7 +(4.2\% of the value))ml
    \item the measured PEEP accuracy  $\pm$(0.1 +(0.3\% of the value))\si{\cmw}
\end{itemize}
It should be noted that the accuracies extracted from comparison with the simulator are overestimates since they neglect the accuracy of the simulator devices which has certainly a finite value. 






%% file: figures/ISO/figurelatex/latex_test_ISO.tex
      \begin{figure}[ht!]
      \centering
       \begin{subfigure}[t]{0.49\linewidth}
        \includegraphics[width=\linewidth]{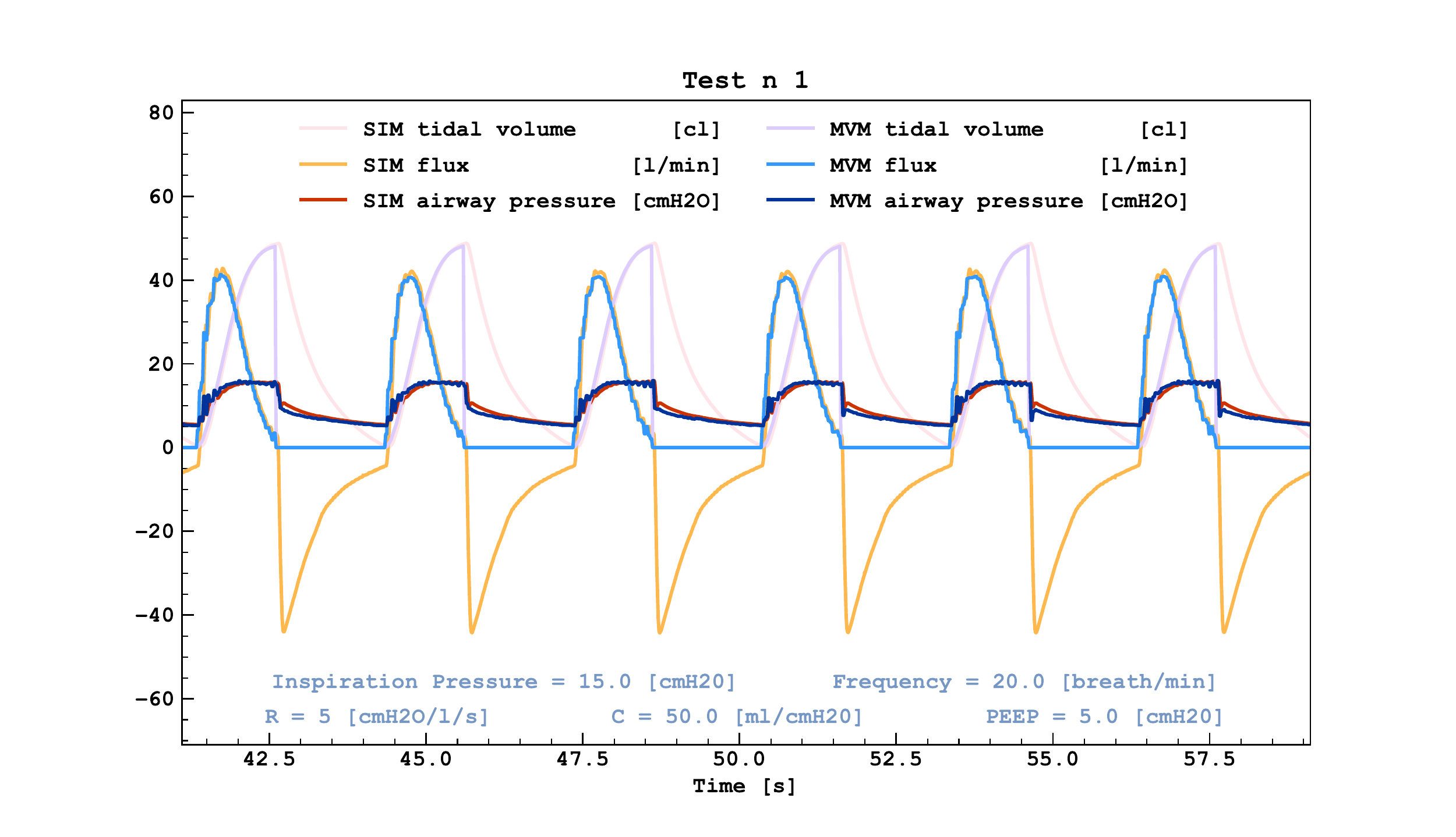}
      \caption{
TV=500, C=50, R=5, p=15,  r=20, PEEP=5}
      \end{subfigure} \begin{subfigure}[t]{0.49\linewidth}
        \includegraphics[width=\linewidth]{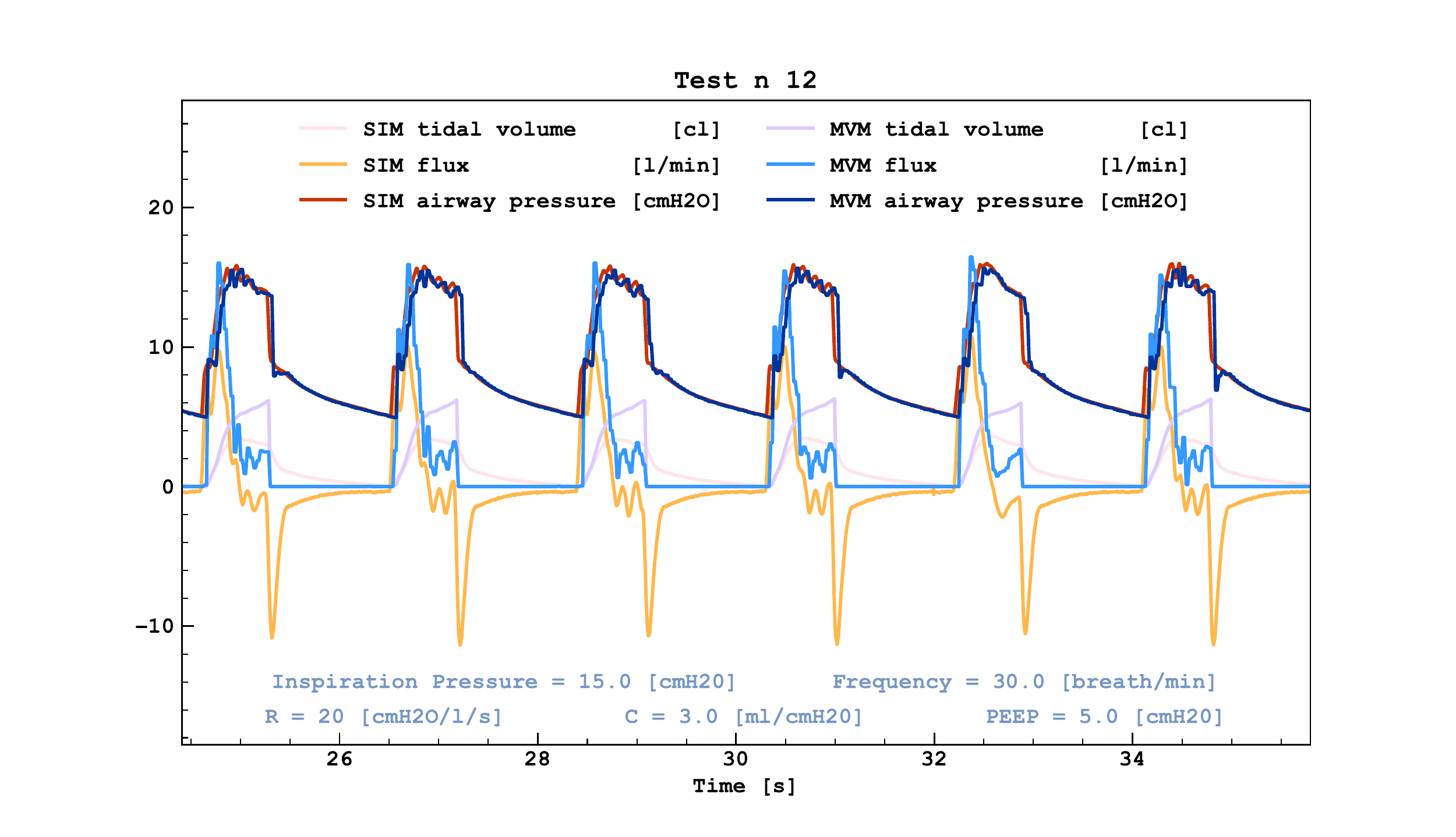}
      \caption{
TV=30, C=3, R=20, p=15,  r=30, PEEP=5}
      \end{subfigure}
       \begin{subfigure}[t]{0.49\linewidth}
        \includegraphics[width=\linewidth]{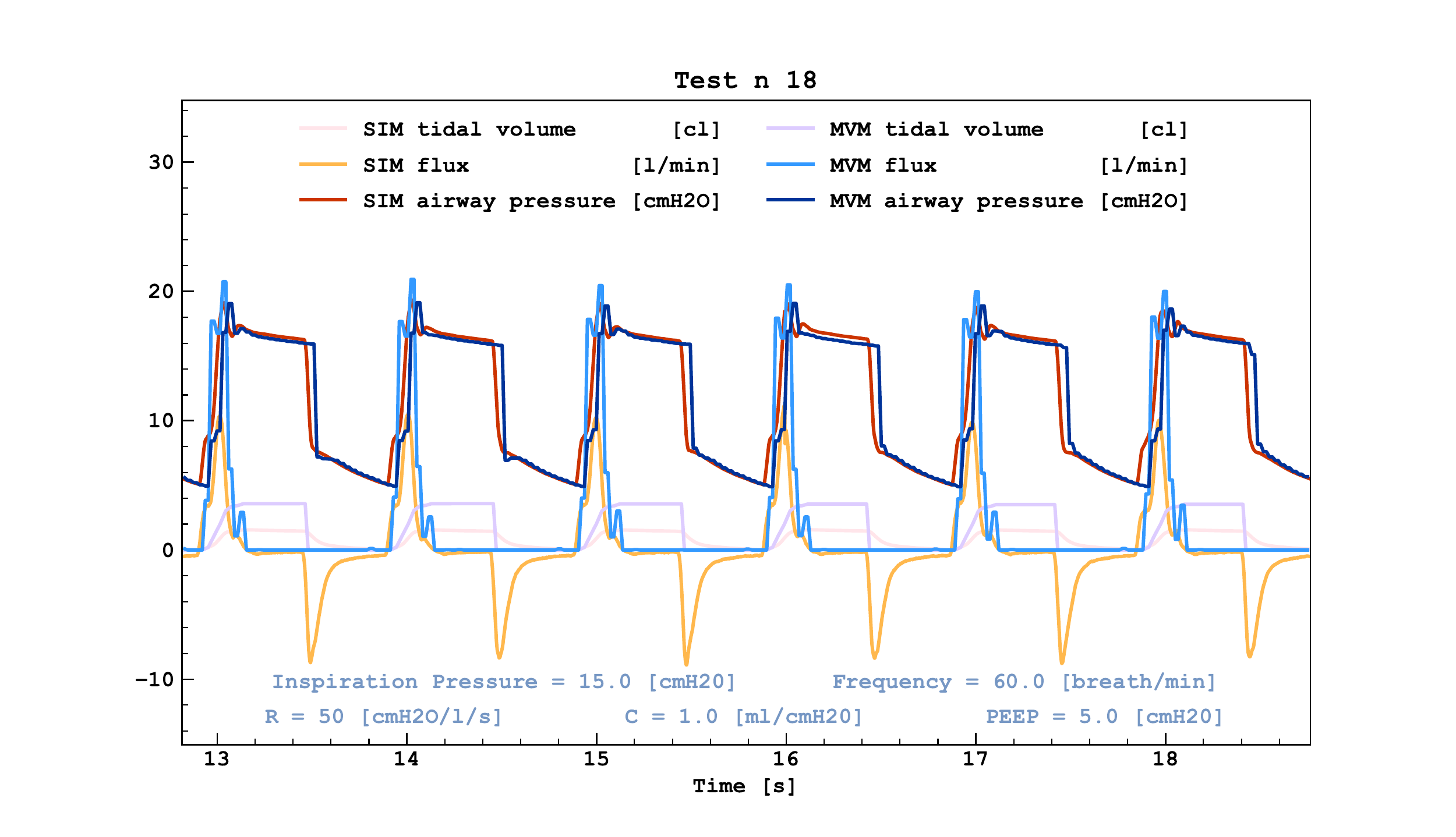}
      \caption{
TV=10, C=1, R=50, p=15,  r=60, PEEP=5}
      \end{subfigure} \begin{subfigure}[t]{0.49\linewidth}
        \includegraphics[width=\linewidth]{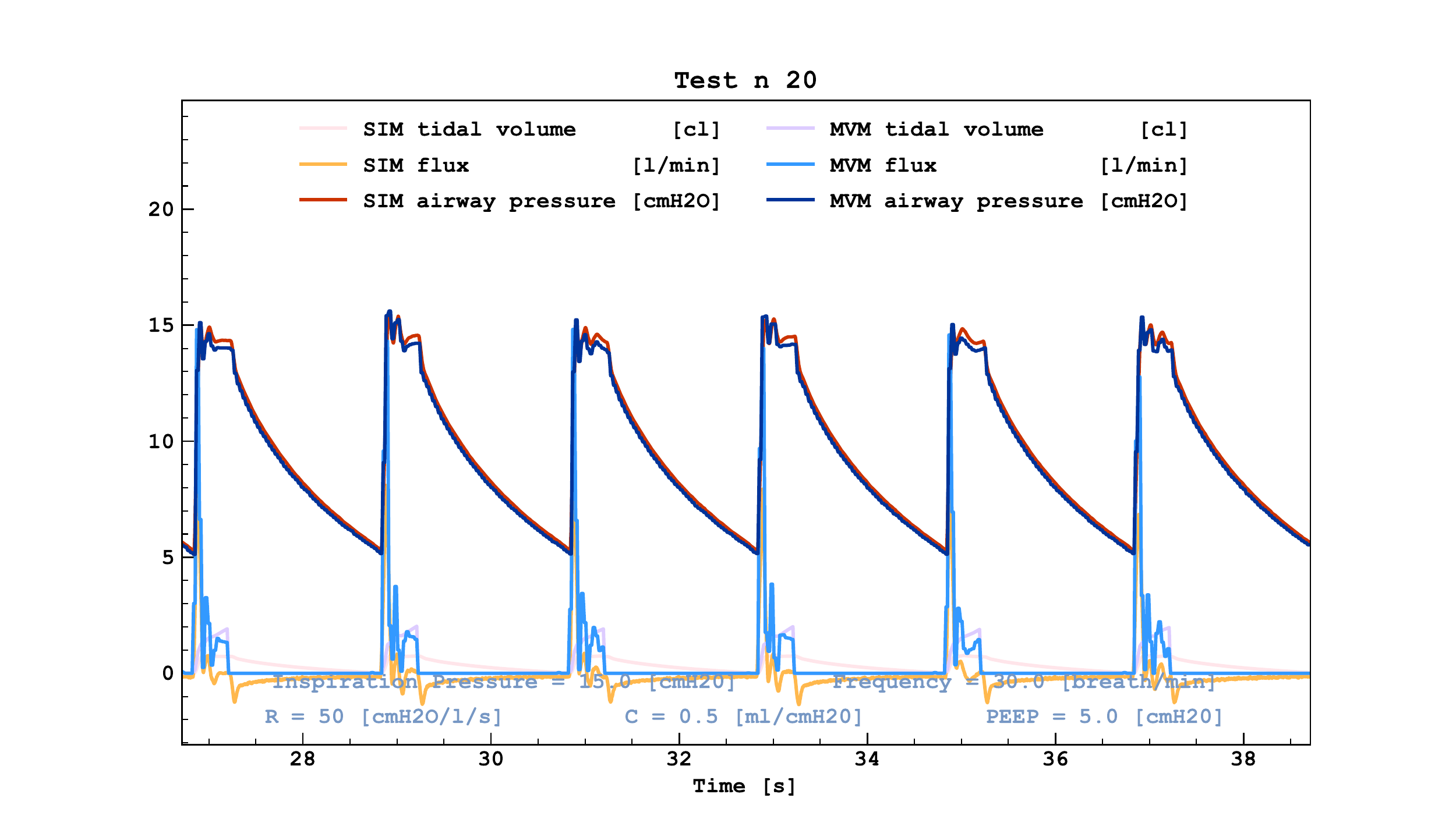}
      \caption{
TV=5, C=0.5, R=50, p=15, r=30, PEEP=5}
      \end{subfigure}
       \begin{subfigure}[t]{0.49\linewidth}
        \includegraphics[width=\linewidth]{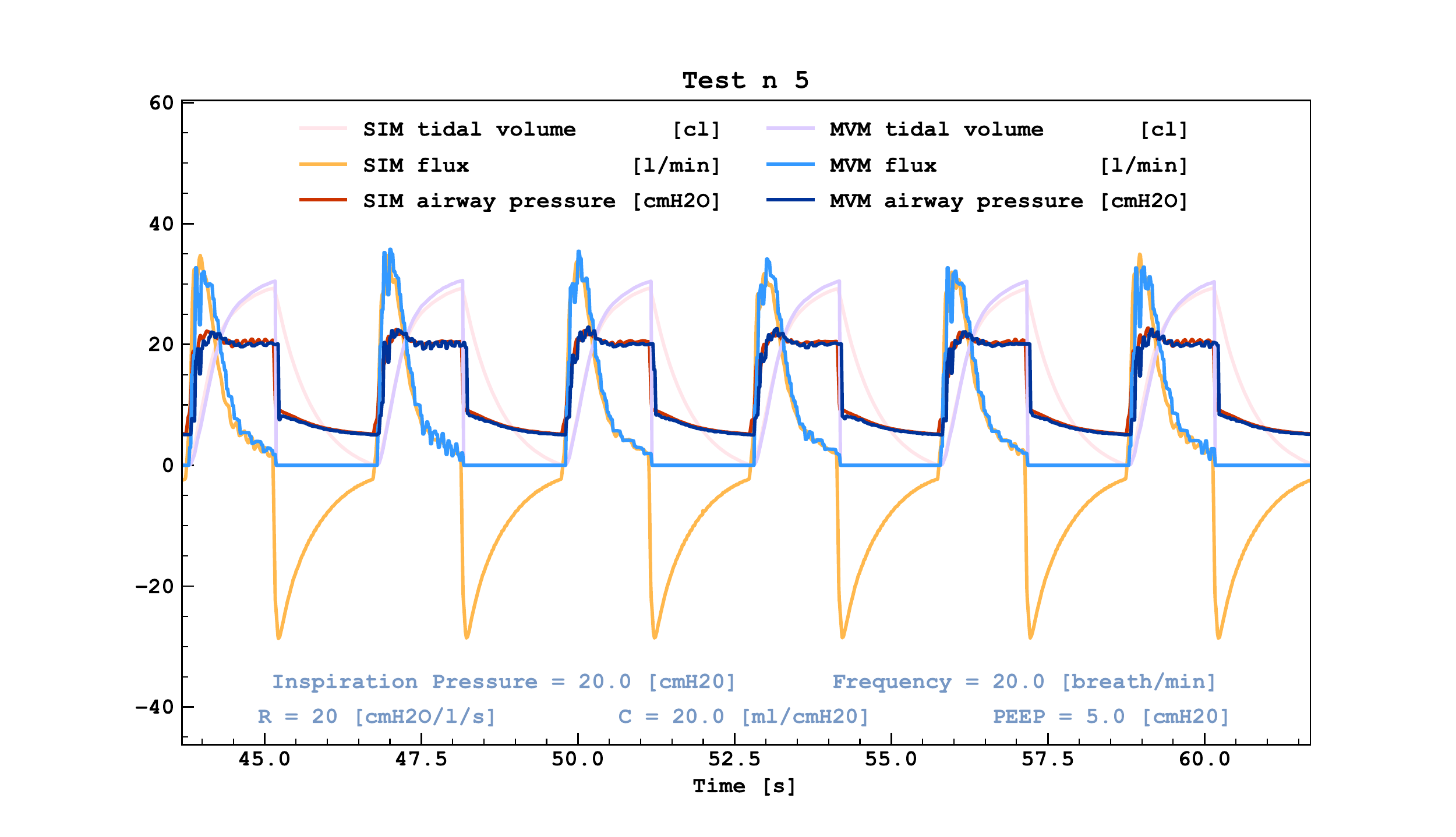}
      \caption{
TV=300, C=20, R=20, p=20, r=20, PEEP=5}
      \end{subfigure} \begin{subfigure}[t]{0.49\linewidth}
        \includegraphics[width=\linewidth]{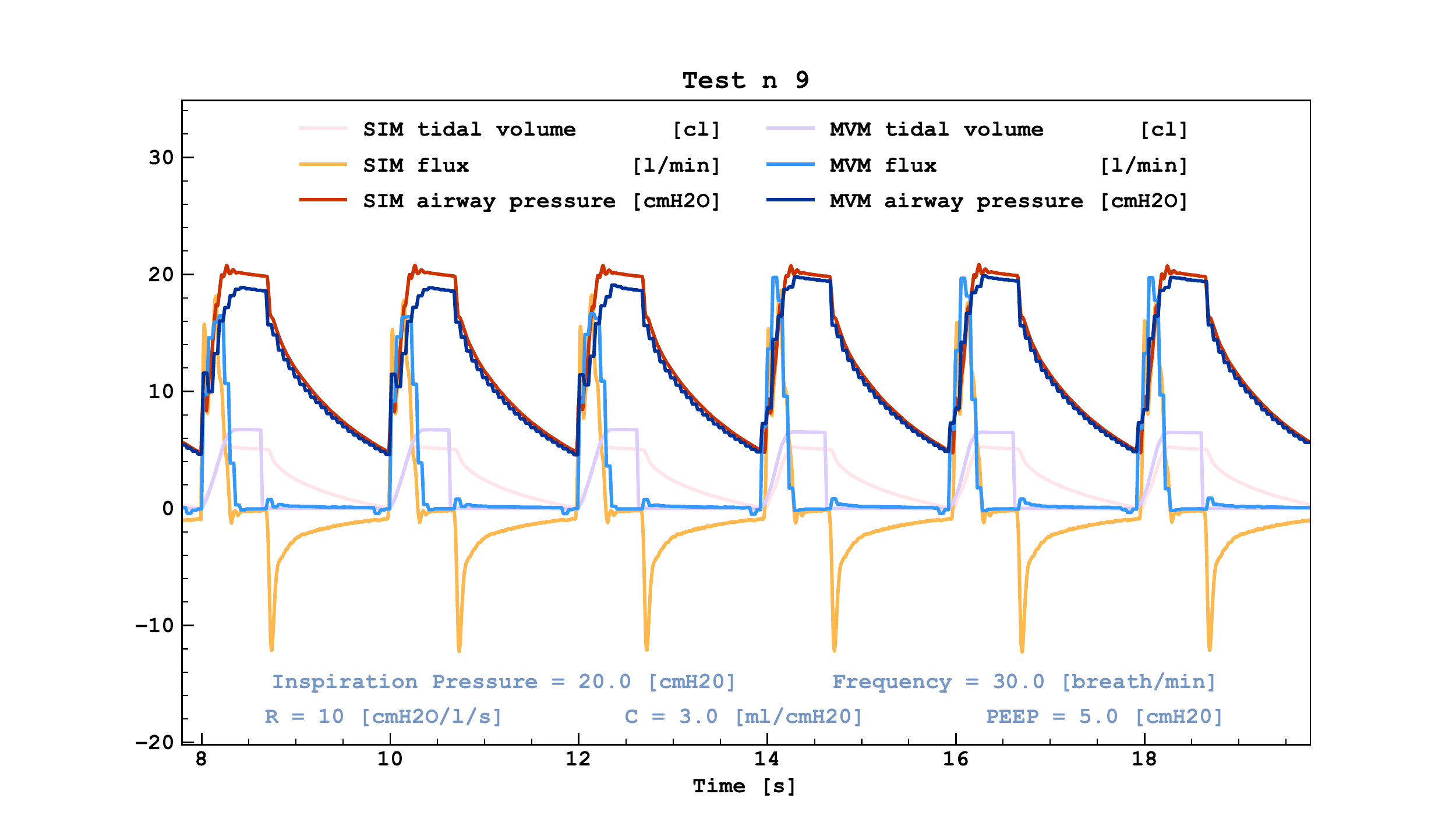}
      \caption{
TV=50, C=10, R=10, p=20,  r=30, PEEP=5}
      \end{subfigure} 

      \caption{The ISO test results with fixed \FiOTwo\ at \SI{21}{\%} are shown for each configuration: intended tidal volume (TV) in \si{ml}, compliance (C) in \si{ml/\cmw}, resistance (R) in \si{\cmw/l/s}, inspiratory pressure (p) in \si{\cmw}, rate (r) in breaths/min, and \PEEP\ in \si{\cmw}.  The test number on top of the each plot corresponds to the test number in the ISO standard.}
      \label{fig:ISOTest-1}

        \end{figure}

      \begin{figure}[ht!]
      \centering
       \begin{subfigure}[t]{0.49\linewidth}
        \includegraphics[width=\linewidth]{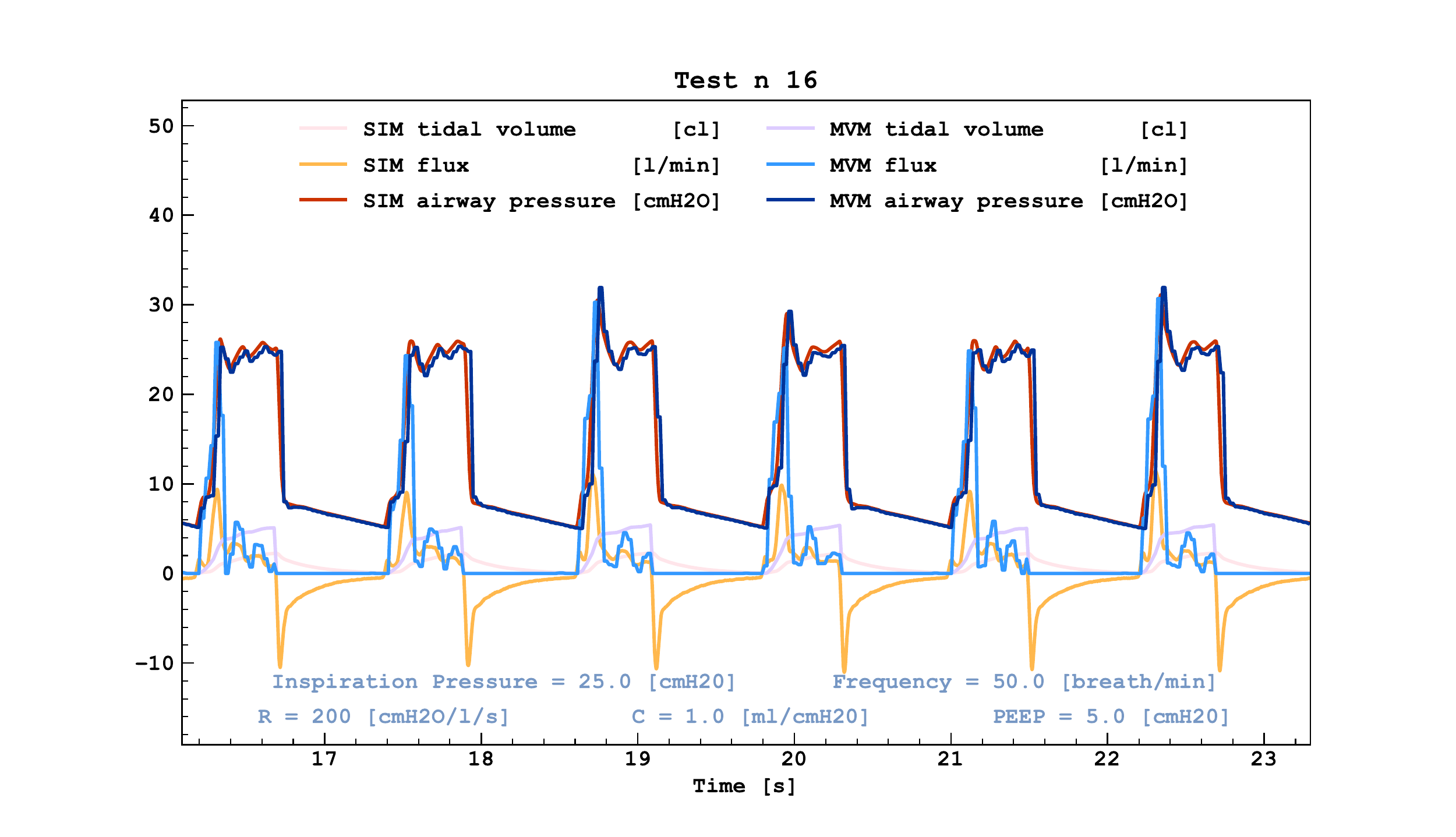}
      \caption{
TV=20, C=1, R=200, p=25,  r=50, PEEP=5}
      \end{subfigure} \begin{subfigure}[t]{0.49\linewidth}
        \includegraphics[width=\linewidth]{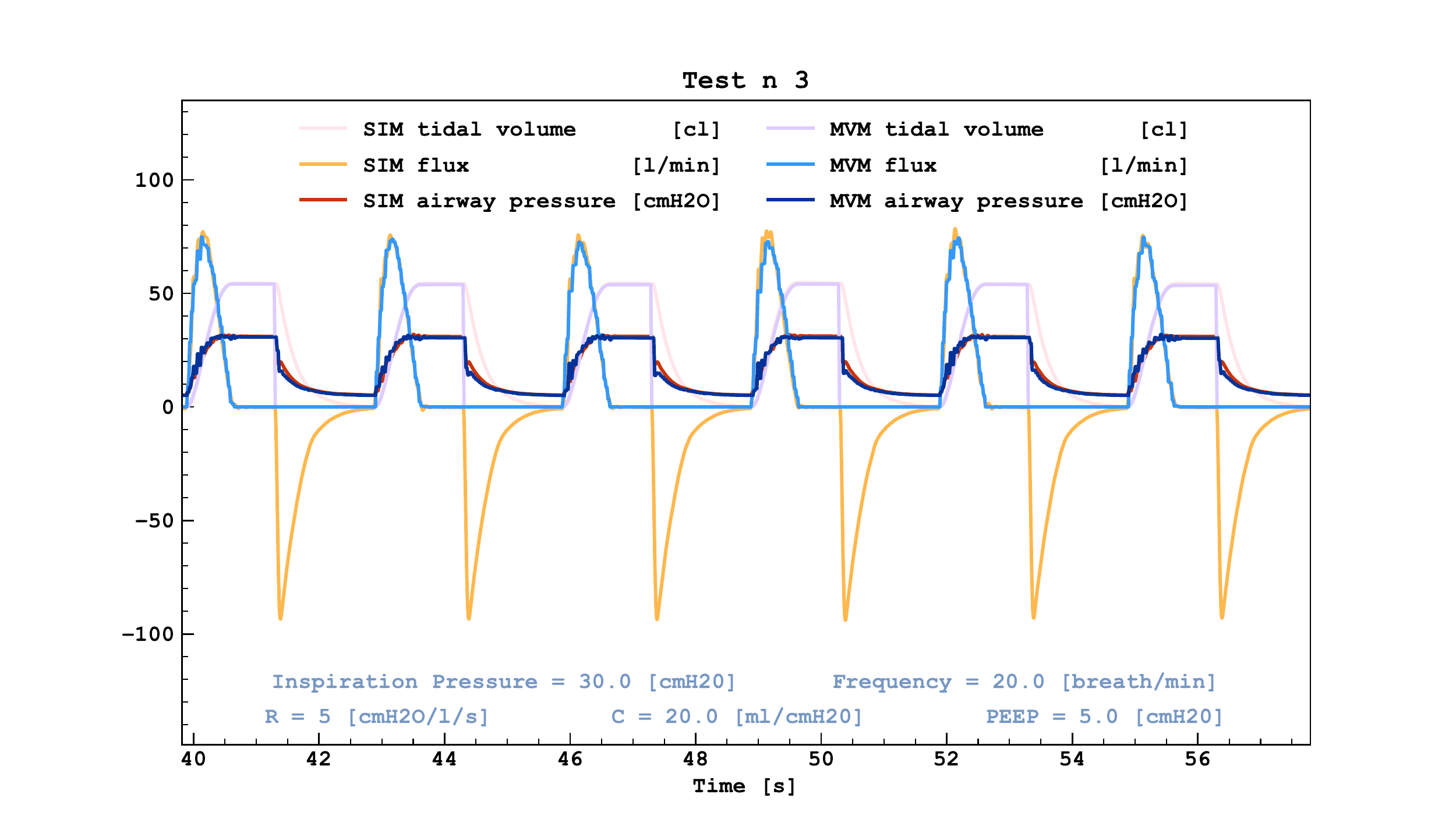}
      \caption{
TV=500, C=20, R=5, p=30,  r=20, PEEP=5}
      \end{subfigure}
       \begin{subfigure}[t]{0.49\linewidth}
        \includegraphics[width=\linewidth]{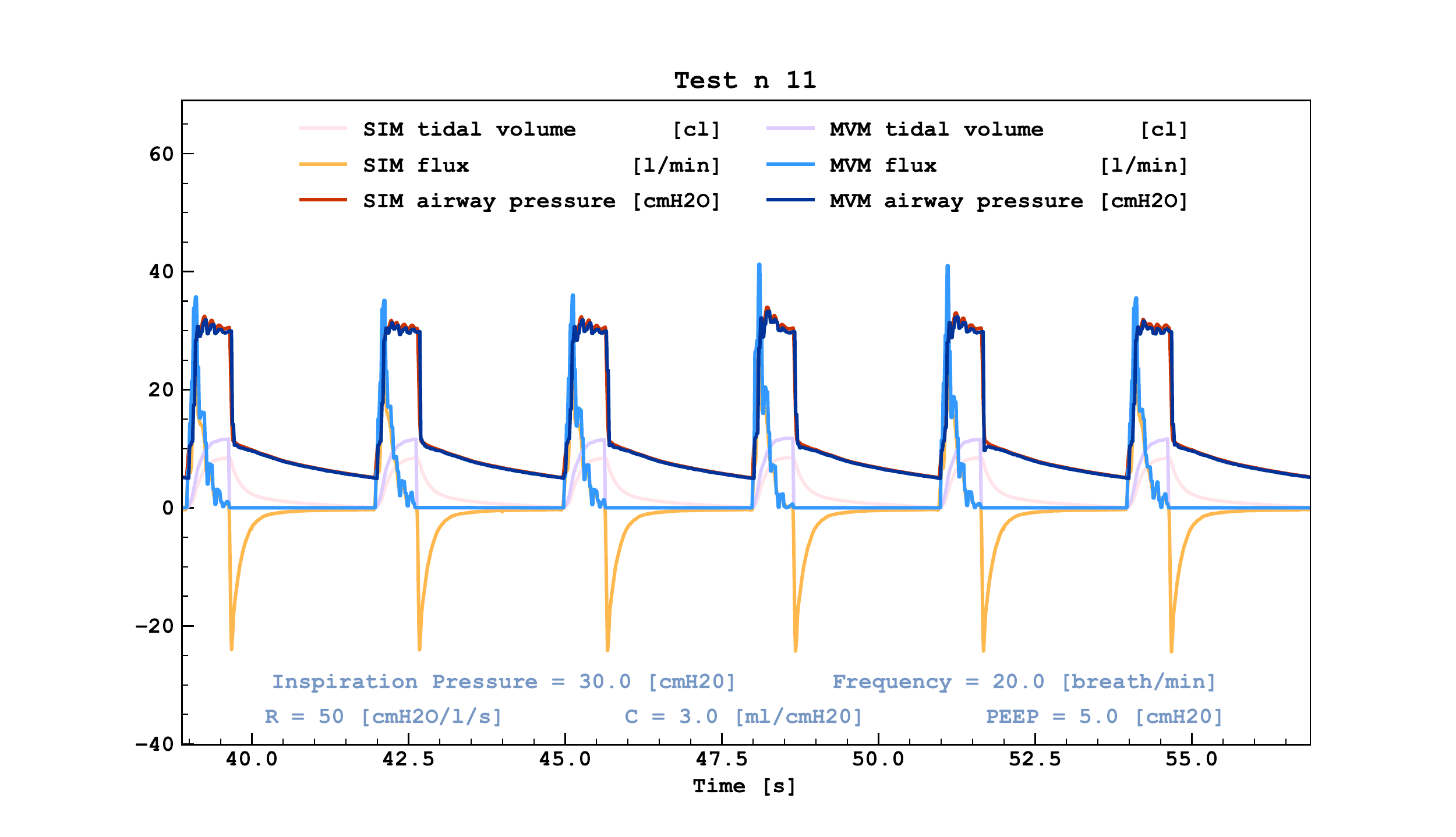}
      \caption{
TV=50, C=3, R=50, p=30,  r=20, PEEP=5}
      \end{subfigure} \begin{subfigure}[t]{0.49\linewidth}
        \includegraphics[width=\linewidth]{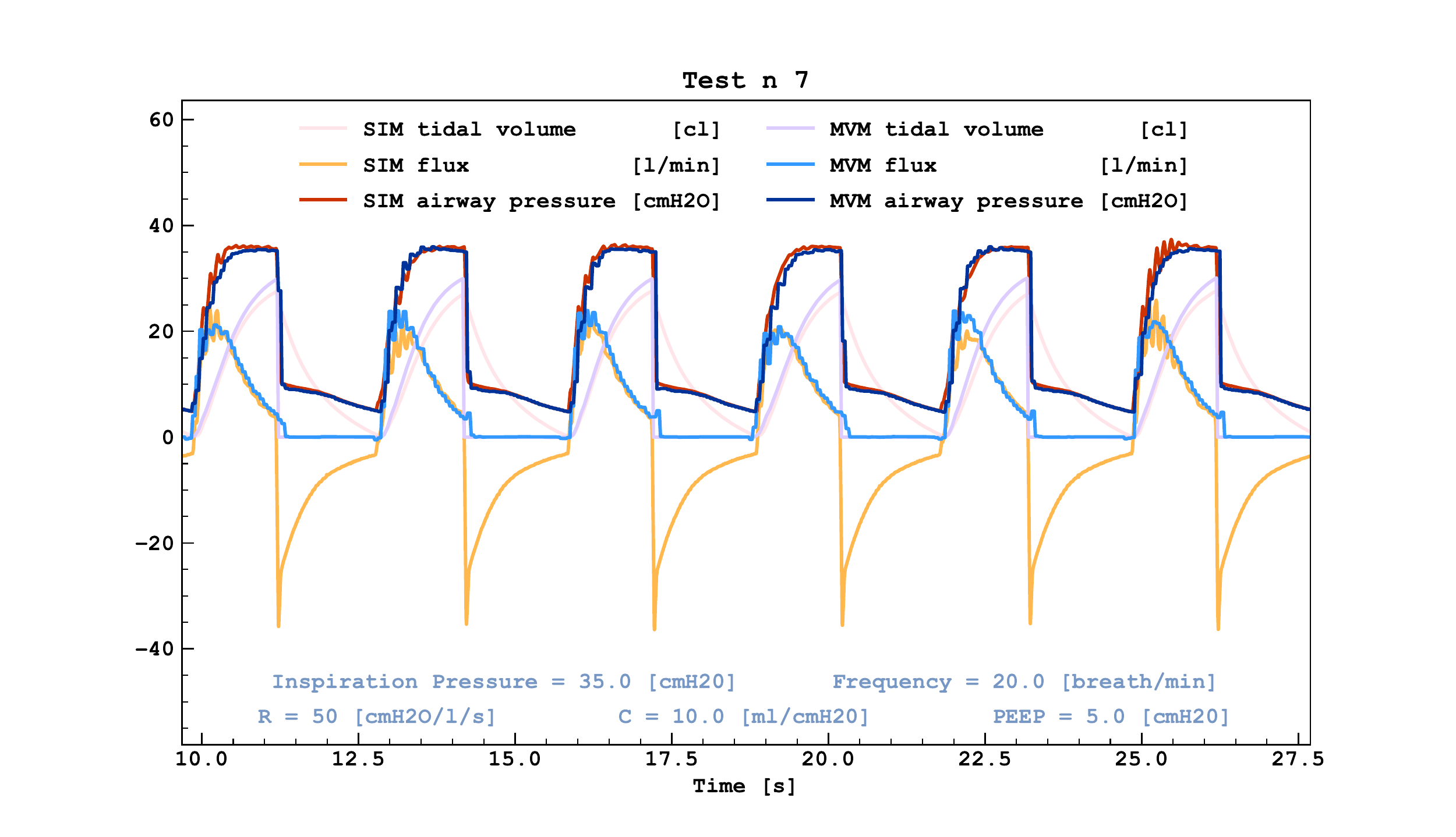}
      \caption{
TV=300, C=50, R=50, p=35, r=20, PEEP=5}
      \end{subfigure}
       \begin{subfigure}[t]{0.49\linewidth}
        \includegraphics[width=\linewidth]{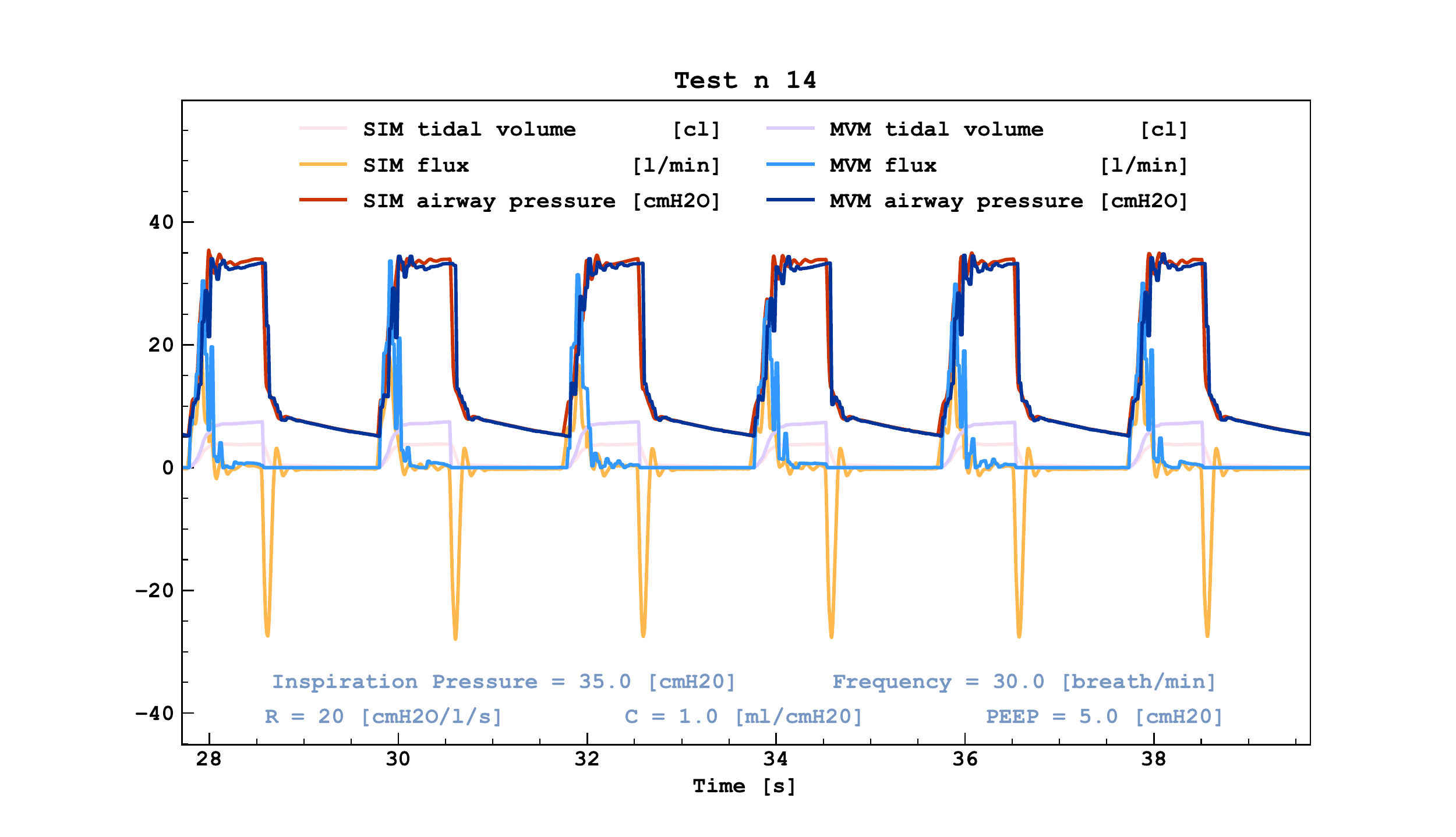}
      \caption{
TV=30, C=1, R=20, p=35, r=30, PEEP=5}
      \end{subfigure} \begin{subfigure}[t]{0.49\linewidth}
        \includegraphics[width=\linewidth]{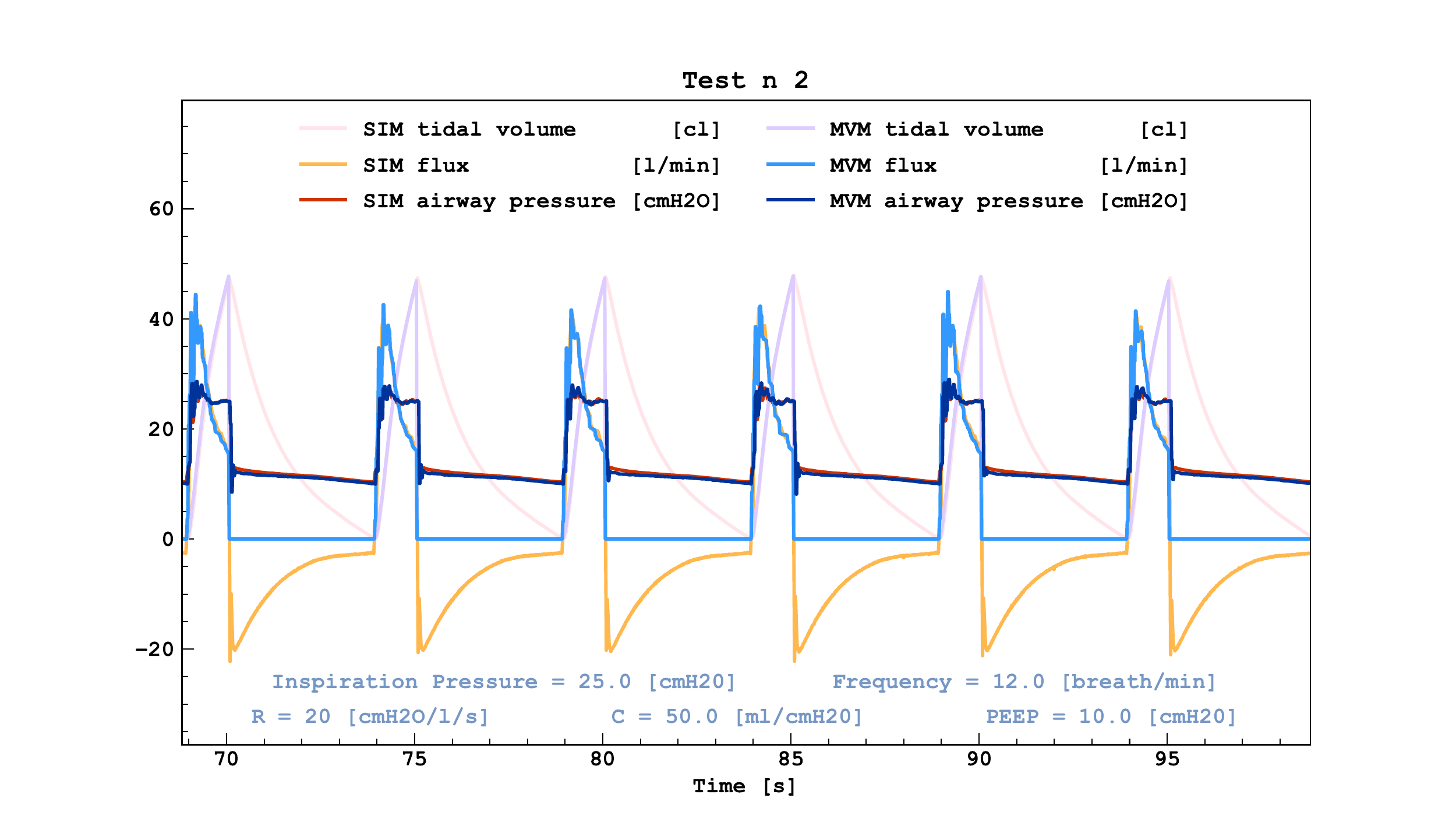}
      \caption{
TV=500, C=50, R=20, p=25, r=12, PEEP=10}
      \end{subfigure}

       \begin{subfigure}[t]{0.49\linewidth}
        \includegraphics[width=\linewidth]{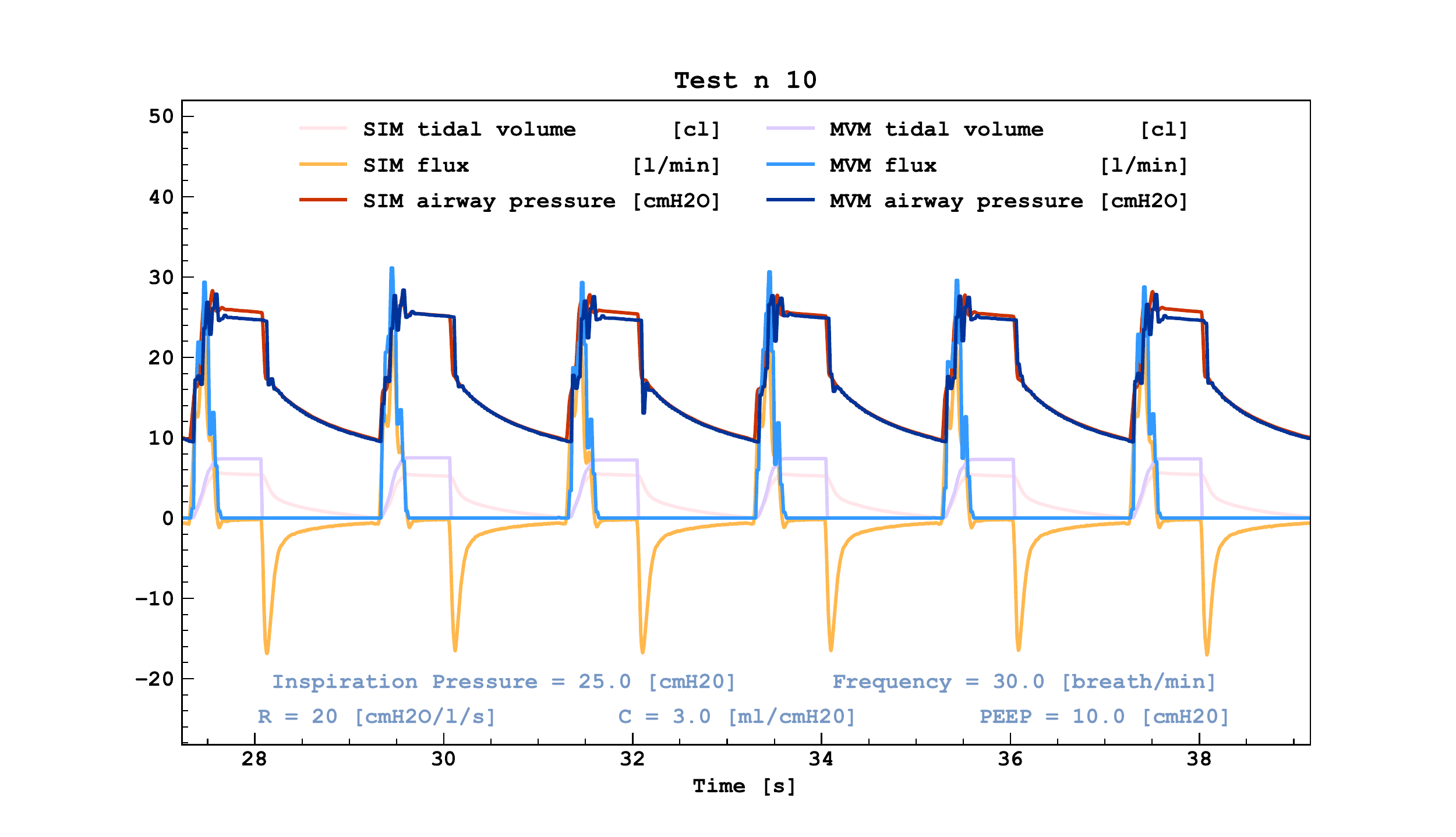}
      \caption{
TV=50, C=3, R=20, p=25, r=30, PEEP=10}
      \end{subfigure} \begin{subfigure}[t]{0.49\linewidth}
        \includegraphics[width=\linewidth]{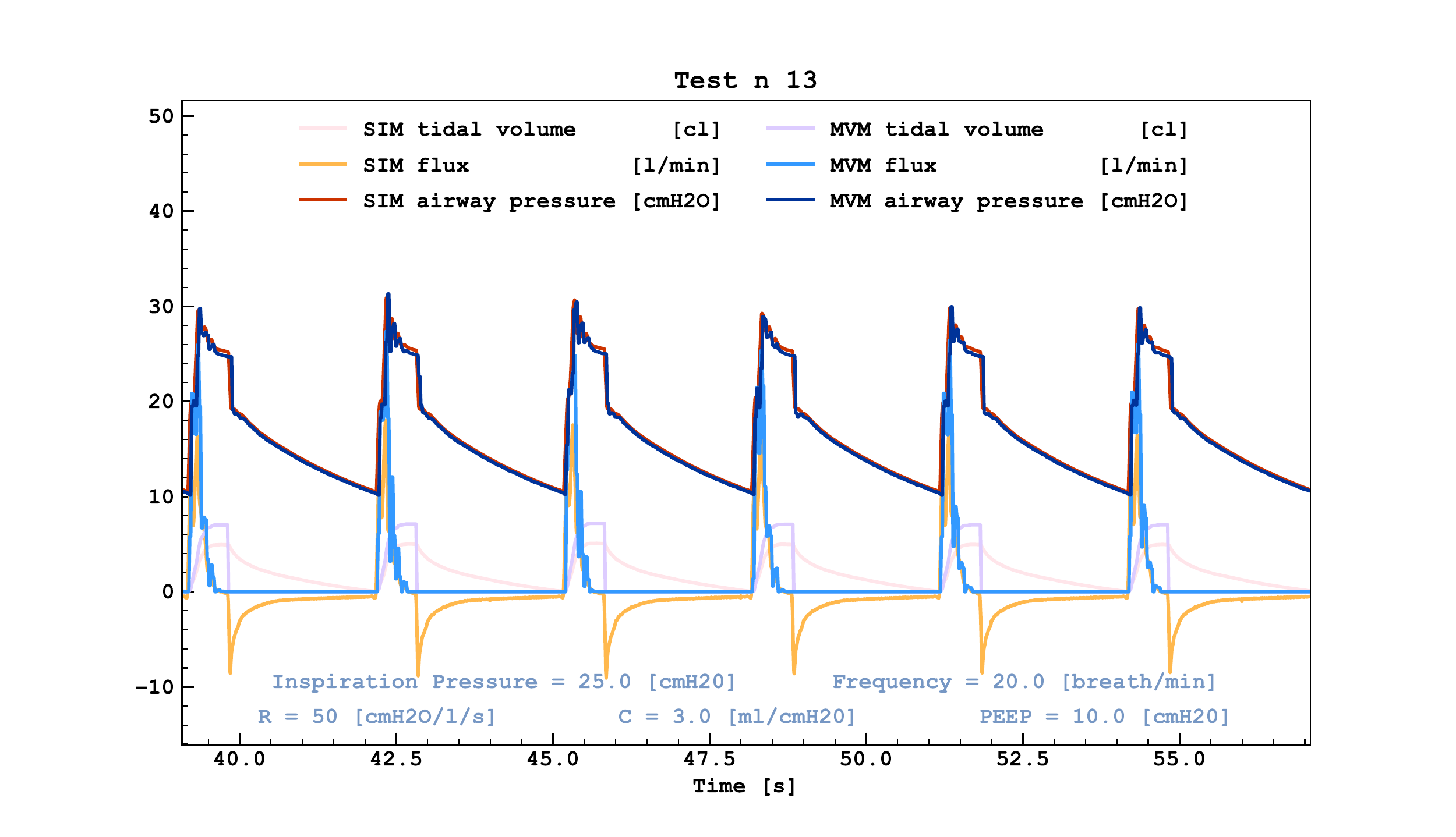}
      \caption{
TV=30, C=3, R=50, p=25, r=20, PEEP=10}
      \end{subfigure}
      
            \caption{The ISO test results with fixed \FiOTwo\ at \SI{21}{\%} are shown for each configuration: intended tidal volume (TV) in \si{ml}, compliance (C) in \si{ml/\cmw}, resistance (R) in \si{\cmw/l/s}, inspiratory pressure (p) in \si{\cmw}, rate (r) in breaths/min, and \PEEP\ in \si{\cmw}.  The test number on top of the each plot corresponds to the test number in the ISO standard.}
      \label{fig:ISOTest-2}

        \end{figure}

      \begin{figure}[ht!]
      \centering
       \begin{subfigure}[t]{0.49\linewidth}
        \includegraphics[width=\linewidth]{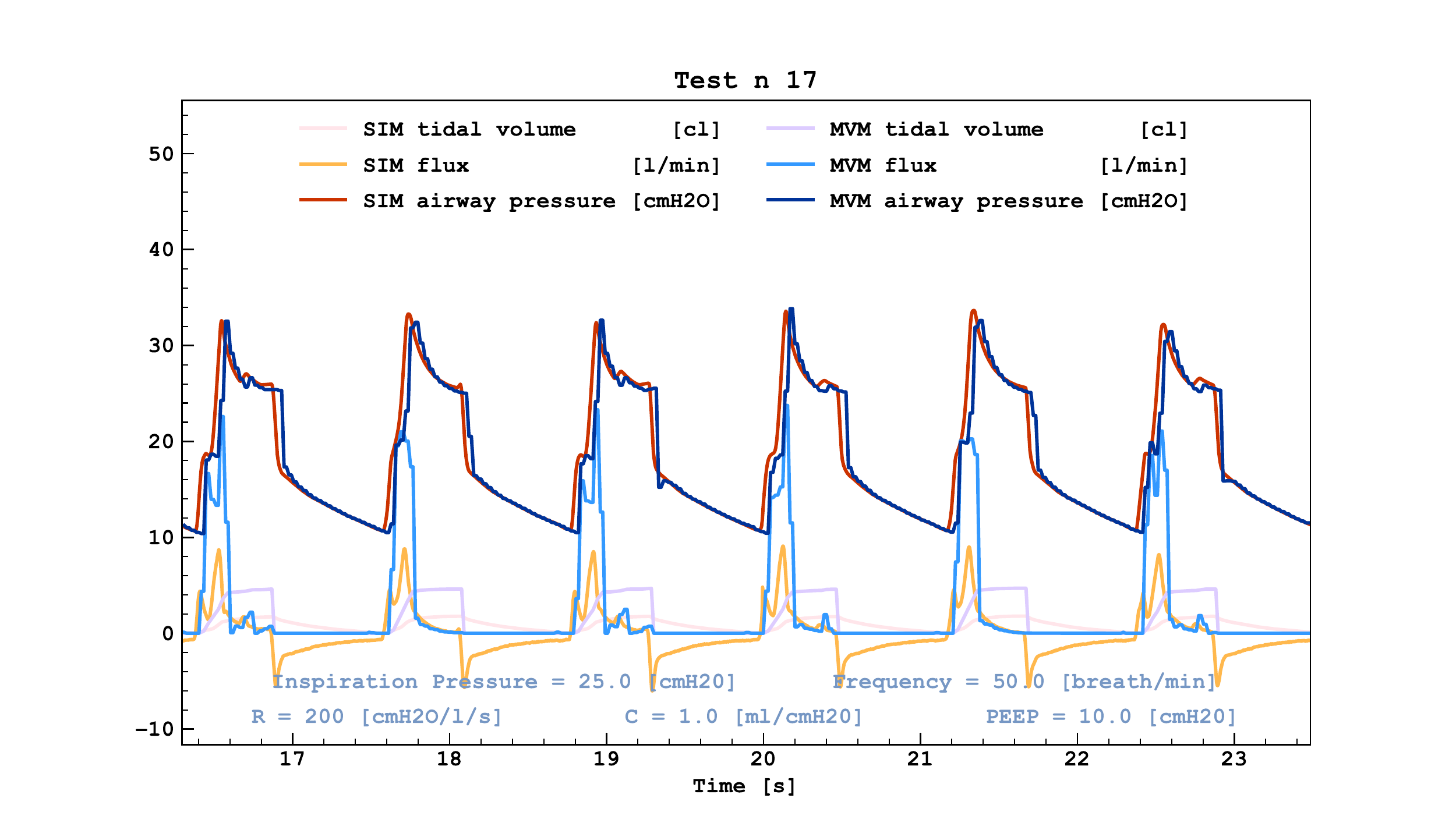}
      \caption{
TV=15, C=1, R=200, p=25, r=50, PEEP=10}
      \end{subfigure} \begin{subfigure}[t]{0.49\linewidth}
        \includegraphics[width=\linewidth]{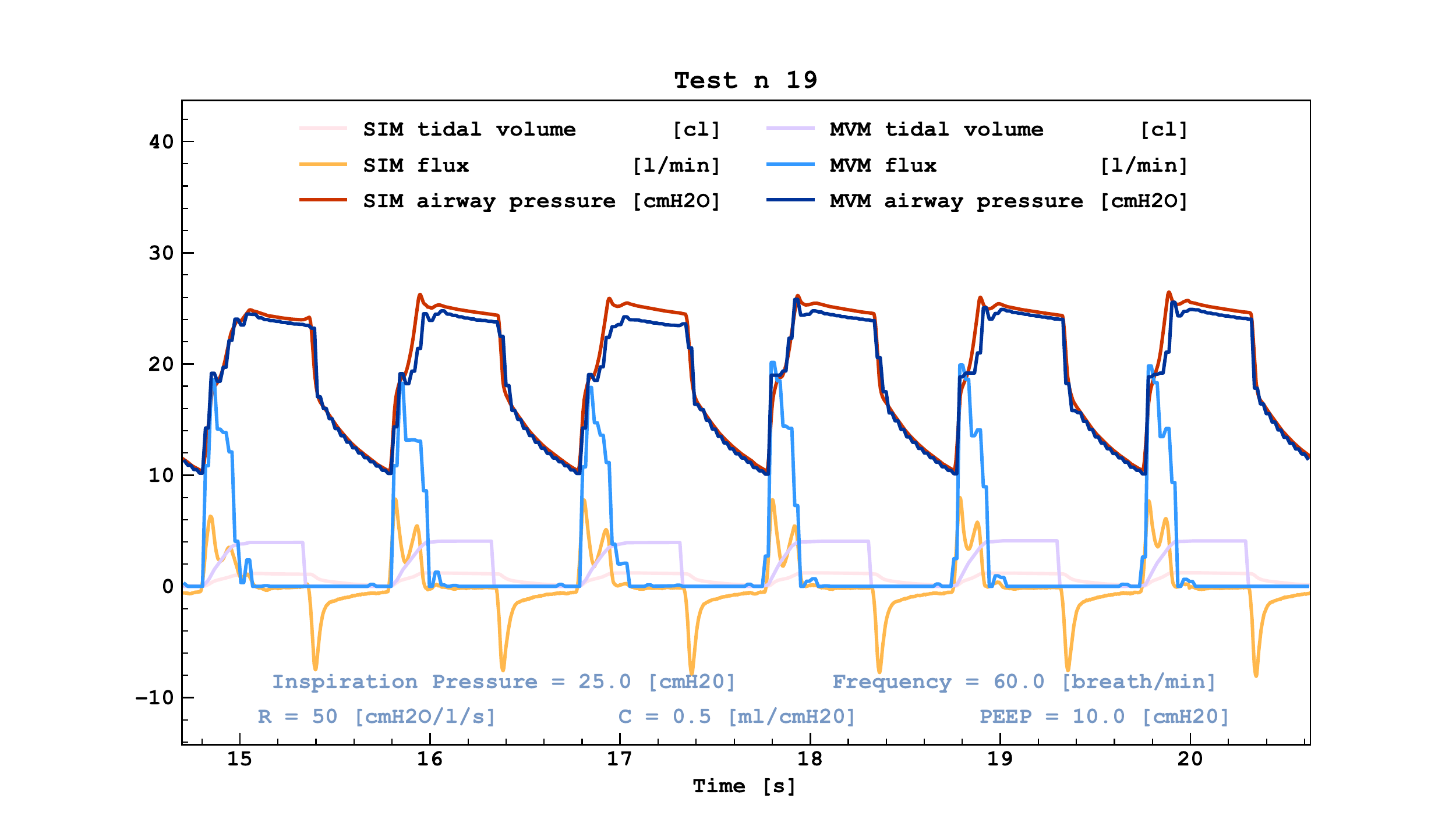}
      \caption{
TV=5, C=0.5, R=50, p=25,  r=60, PEEP=10}
      \end{subfigure}
       \begin{subfigure}[t]{0.49\linewidth}
        \includegraphics[width=\linewidth]{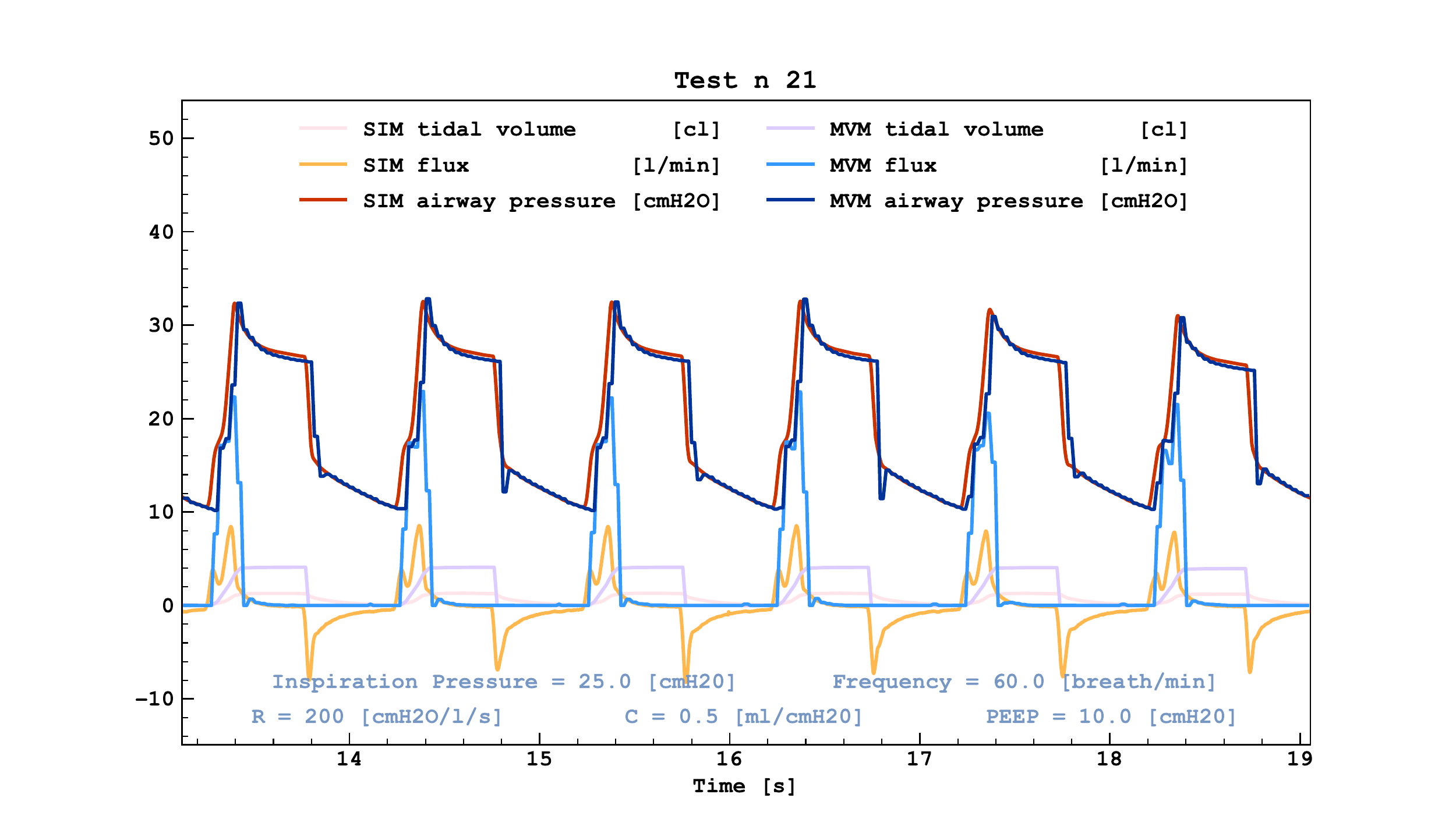}
      \caption{
TV=5, C=0.5, R=200, p=25, r=60, PEEP=10}
      \end{subfigure} \begin{subfigure}[t]{0.49\linewidth}
        \includegraphics[width=\linewidth]{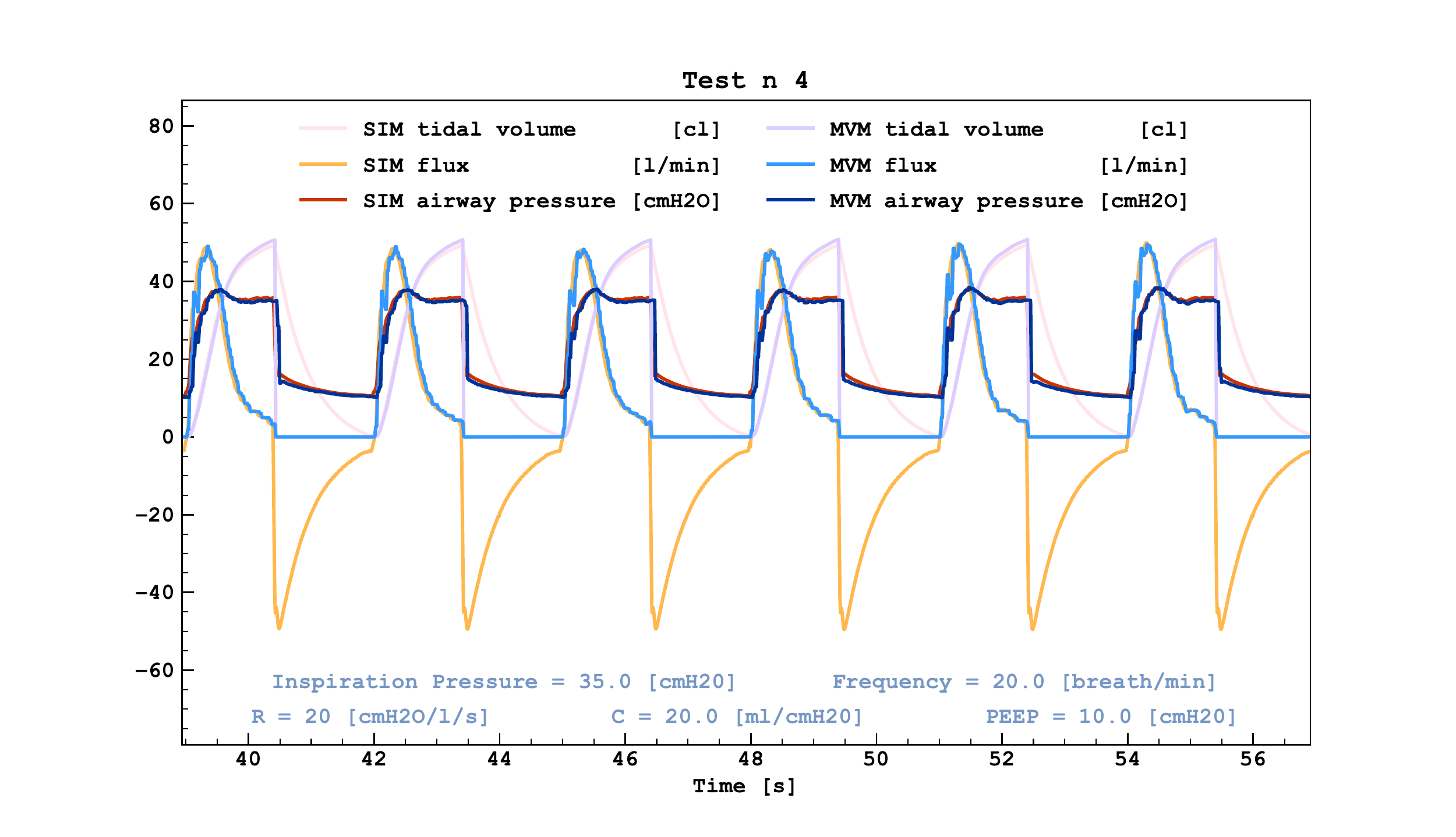}
      \caption{
TV=500, C=20, R=20, p=35,  r=20, PEEP=10}
      \end{subfigure} 

       \begin{subfigure}[t]{0.49\linewidth}
        \includegraphics[width=\linewidth]{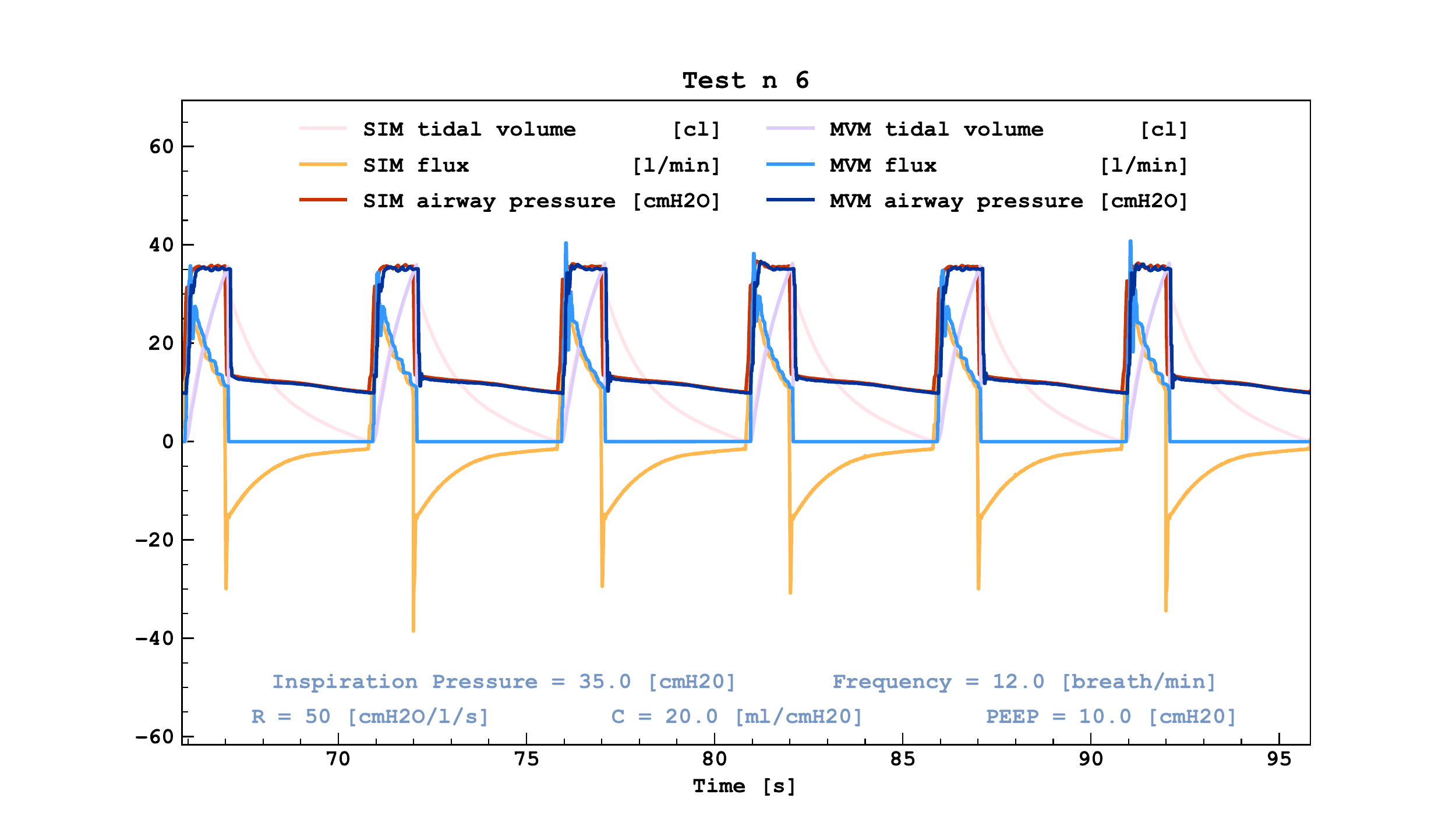}
      \caption{
TV=300, C=20, R=50, p=35, r=12, PEEP=10}
      \end{subfigure} \begin{subfigure}[t]{0.49\linewidth}
        \includegraphics[width=\linewidth]{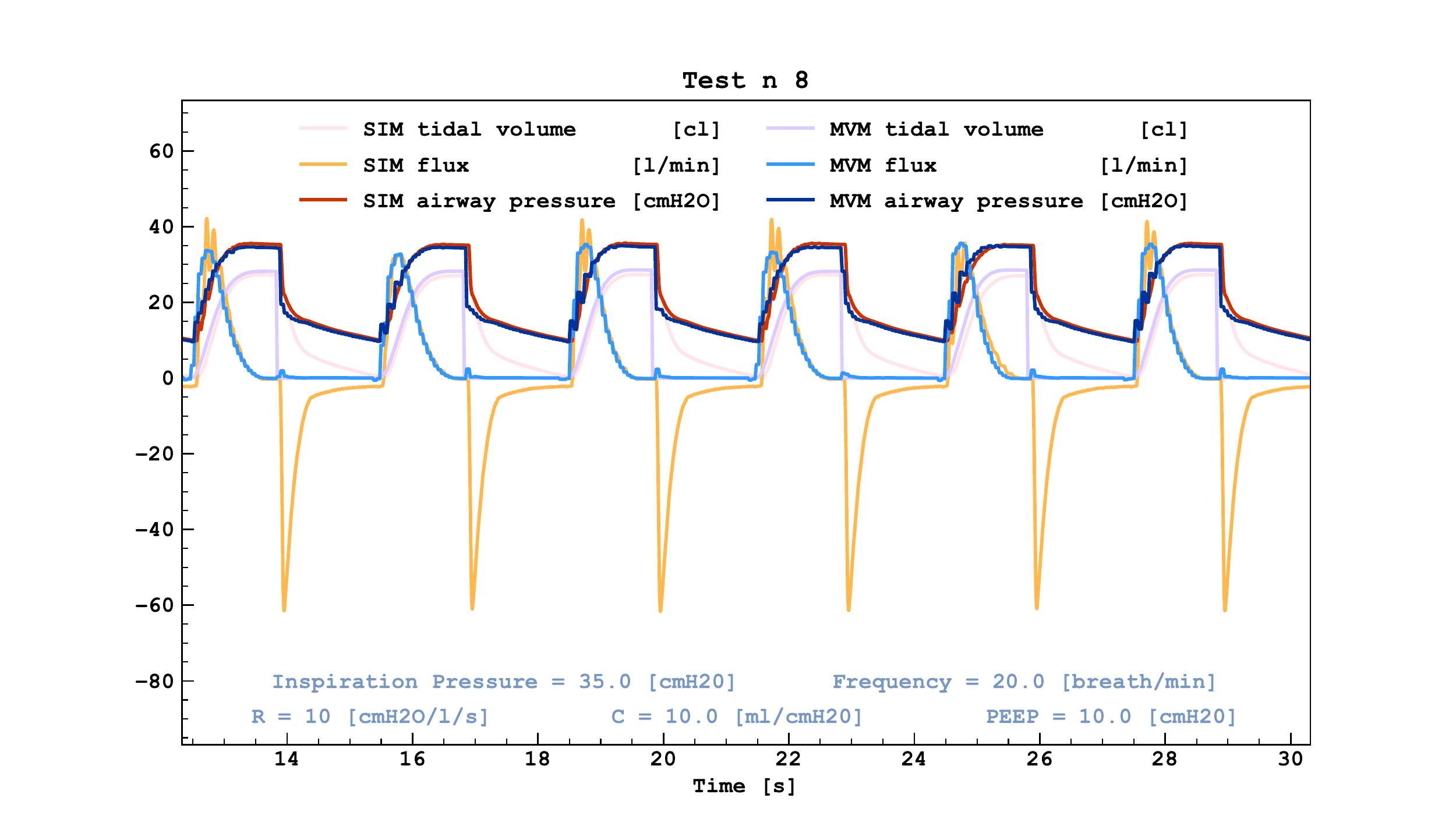}
      \caption{
TV=200, C=10, R=10, p=35, r=20, PEEP=10}
      \end{subfigure}

       \begin{subfigure}[t]{0.49\linewidth}
        \includegraphics[width=\linewidth]{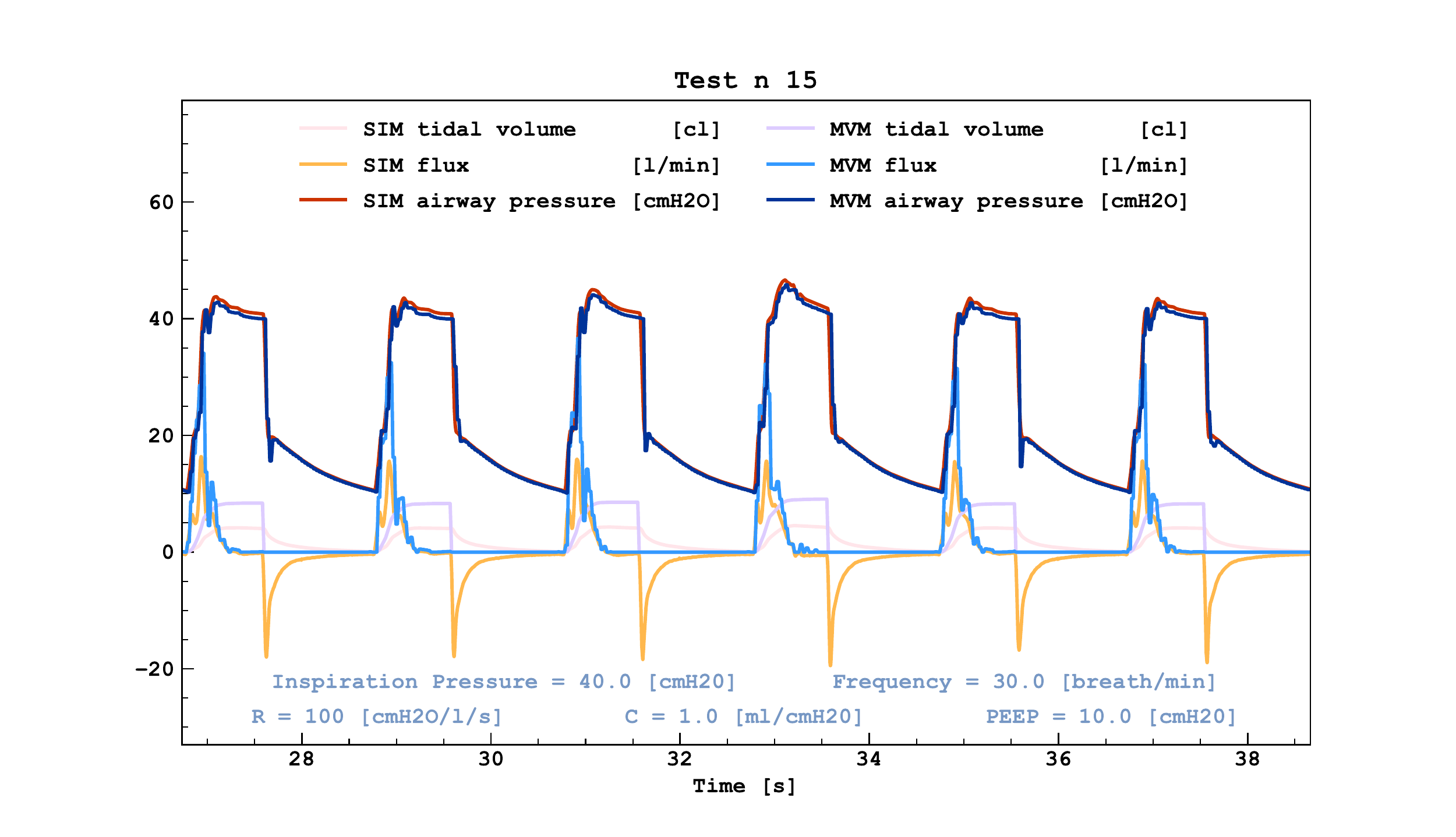}
      \caption{
TV=30, C=1, R=100, p=40,  r=30, PEEP=10}
      \end{subfigure}

      \caption{The ISO test results with fixed \FiOTwo\ at \SI{21}{\%} are shown for each configuration: intended tidal volume (TV) in \si{ml}, compliance (C) in \si{ml/\cmw}, resistance (R) in \si{\cmw/l/s}, inspiratory pressure (p) in \si{\cmw}, rate (r) in breaths/min, and \PEEP\ in \si{\cmw}.  The test number on top of the each plot corresponds to the test number in the ISO standard.}
      \label{fig:ISOTest-4}

    \end{figure}

%% file: sections/Assisted.tex
\section{Tests in pressure-supported ventilation  mode}
\label{sec:assisted}

\begin{figure*}[t!]
\centering{}
\includegraphics[width=0.49\textwidth]{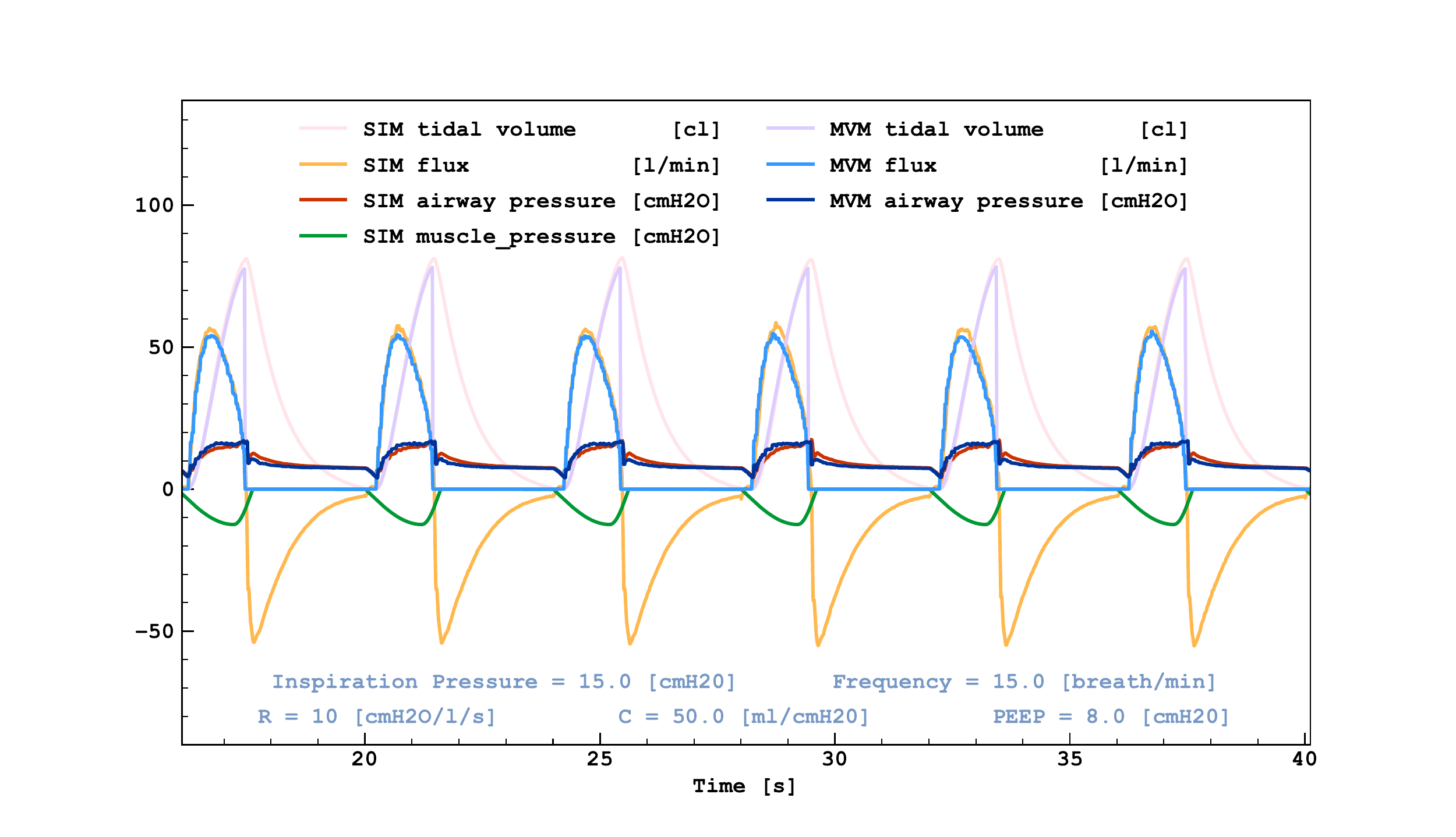}
\includegraphics[width=0.49\textwidth]{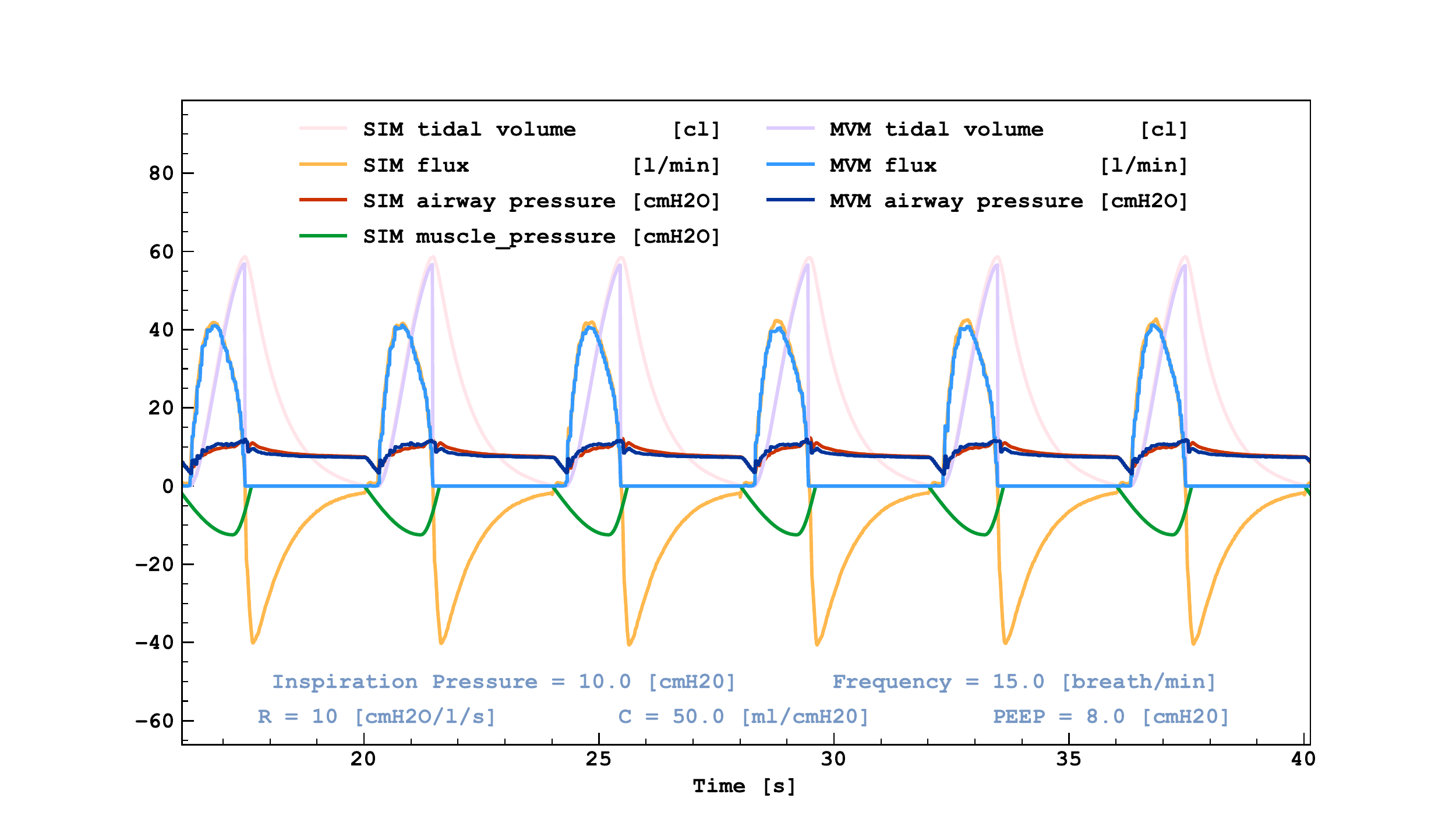}
\caption[\MVM\ Illustration.]{\label{fig:MVM-assisted} Waveforms with pressure support ventilation  mode. The green line represents the muscle pressure as exerted by the breathing simulator.}
\end{figure*}

Waveforms with pressure-supported ventilation  mode are shown in \reffig{MVM-assisted} for one set of parameters. The patient is this case initiates the breathing and triggers the pressure support.

%% file: sections/addtional.tex
\section{New features under  implementation}
\label{sec:additional}

The MVM will also integrate advanced features designed by anaesthesiologists participating to the project who happen to work in the medical wards in Lombardy, the region most severely hit by the COVID-19 epidemics.  MVM will enable at the touch of a button the measurements of two vital parameters crucial for the determination of the best course of care for COVID-19 patients.

The first parameter is the Plateau Pressure (PP), the pressure reached inside the alveoli at the end of the inspiratory cycle.  PP may be lower than the Set Inspiratory Pressure (SIP) provided by the ventilator.  The difference between PP and the PEEP is called Driving Pressure, DP (DP = PP – \PEEP).  The PP is measured through a forced hold at the end of inspiration, activated through the Inspiratory Hold Maneuver (IHM) button through the touch-screen of the GUI.  When the IHM is pressed, the MVM will wait for the end of the next inspiration phase, and if the IHM is still pressed, at that moment will hold both the inspiratory and expiratory valves closed till the IHM button is released.  An emergency reset button allows to override internal controls and to resume the regular breathing cycle.

The PP is the pressure reached in the airway at the end of the IHM.  At the end of IHM,
the screen will show a freeze frame of the cycle and will show the number of the measured PP.

The DP should be kept below \SI{14}{\cmw}, as a pressure value too high may results in long term damage to the lungs. Measurement of PP and DP should be performed early in the start of care of sedated patients, to fine tune the best care regimen, which typically start by setting a Tidal Volume (TV) of 6-8 ml/kg/IBW (Ideal Body Weight).  The DP permits to fine tune the TV delivered to the patient, taking note of the portion of the lung under viral attack and permitting doctors to determine the best adjustment of the TV delivered to the patients. 

The second parameter is the AutoPEEP, which may be zero for most patients or significantly different from zero for patients that have obstructions in the exhalation channel, as possibly generated by secretions.  In this case, the small flow during exhalation may result in an incomplete drain of the alveoli during the expiration phase.  An expiratory hold maneuver permits to momentarily close both inspiratory and expiratory valve at the end of the expiration phase, and measure the residual pressure in the alveoli above he \PEEP\ level, the residual value being \PEEP\ + AutoPEEP.  This measurement is once again performed by pushing a single button.  At the press of the button, the expiratory hold will be performed at the end of the following expiratory phase, and the hold action will be completed at the release of the button.  At this instant, the screen will show a freeze frame of the cycle and will show the number of the measured AutoPEEP.

The true value of DP is (PP – \PEEP\ – AutoPEEP).  Measurement of both PP and AutoPEEP is crucial for the determination of DP.

MVM will also carry out at the touch of a button the lung recruitment procedure, i.e., the Recruitment Maneuver (RM), i.e., the emergency procedure required immediately after the end of the intubation.  RM consists in the prolonged lung inflation at increased inspiratory set pressure, as necessary to reactivate the alveoli immediately after intubation.  Before the start of procedure, the doctor must be able to set the Pressure for the Recruitment Maneuver (PRM, from 25 to 50 \si{\cmw}, settable with a $\pm$1 cm pace).  The doctor will also have the choice of setting a fixed time for the RM (Time for Recruitment Maneuver, TRM, settable from 5 to 40 seconds, pace of $\pm$1 sec), with a reset button that can stop the procedure and return the breathing cycle to normal, or will have the choice to keep the RM button pressed and terminate the procedure on release of the button.

%% file: sections/License.tex
\section{License Agreement}
\label{sec:License}

\textcopyright 2020: this paper describes Open Hardware and is licensed under the \MVMLicense.  You may redistribute and modify this documentation and make products using it under the terms of the \MVMLicense~\cite{CERN:2020tw}.\\
This documentation is distributed \textbf{\textit{without any express or implied warranty, including of merchantability, satisfactory quality or fitness for a particular purpose}}.  Please see \MVMLicense~\cite{CERN:2020tw} for applicable conditions.